\documentclass[twocolumn,showkeys]{openjournal}

\usepackage{amsmath}
\usepackage{amssymb}
\usepackage{graphicx}
\usepackage{bm}
\usepackage{url}
\usepackage{hyperref}
\usepackage{xcolor}

\newcommand{\xtrue}{\mathbf{x}_{\rm true}}
\newcommand{\xobs}{\mathbf{x}_{\rm obs}}
\newcommand{\xlat}{\mathbf{x}_{\rm latent}}

\newcommand{\yobs}{\mathbf{y}_{\rm obs}}

\newcommand{\srange}{\sigma_{\rm range}}
\newcommand{\sx}{\sigma_x}
\newcommand{\sy}{\sigma_y}

\begin{document}

\title{Why Machine Learning Models Systematically Underestimate Extreme Values II: \\ How to Fix It with LatentNN \vspace{-1cm}}
\author{Yuan-Sen Ting}
\email{ting.74@osu.edu}
\affiliation{Department of Astronomy, The Ohio State University, Columbus, OH 43210, USA}
\affiliation{Center for Cosmology and AstroParticle Physics (CCAPP), The Ohio State University, Columbus, OH 43210, USA}
\affiliation{Max-Planck-Institut für Astronomie, K\"onigstuhl 17, D-69117 Heidelberg, Germany}

\begin{abstract}
Attenuation bias---the systematic underestimation of regression coefficients due to measurement errors in input variables---affects astronomical data-driven models. For linear regression, this problem was solved by treating the true input values as latent variables to be estimated alongside model parameters. In this paper, we show that neural networks suffer from the same attenuation bias and that the latent variable solution generalizes directly to neural networks. We introduce LatentNN, a method that jointly optimizes network parameters and latent input values by maximizing the joint likelihood of observing both inputs and outputs. We demonstrate the correction on one-dimensional regression, multivariate inputs with correlated features, and stellar spectroscopy applications. LatentNN reduces attenuation bias across a range of signal-to-noise ratios where standard neural networks show large bias. This provides a framework for improved neural network inference in the low signal-to-noise regime characteristic of astronomical data. This bias correction is most effective when measurement errors are less than roughly half the intrinsic data range, in the regime of very low signal-to-noise and few informative features. Code is available at \url{https://github.com/tingyuansen/LatentNN}.
\end{abstract}


\section{Introduction}
\label{sec:intro}

Modern astronomy increasingly relies on neural networks for inference \citep{Cranmer2020a, Huertas-Company2023, Ting2025}. Data-driven approaches now routinely map observed quantities to physical parameters across a wide range of applications, from stellar parameters derived from spectra \citep[e.g.,][]{Ness2015, Fabbro2018, Leung2019, Ting2019} to photometric redshifts from imaging \citep{Hoyle2016, Li2022_GaZNet, Lin2022, Zhou2022} to stellar properties from time series light curves \citep{Pasquet2019, Hon2021}. 

With the rise of neural network approaches, however, input uncertainties are often ignored or left unmodeled. This is largely a design choice rather than an oversight. Classical machine learning methods have long grappled with measurement errors in inputs, developing explicit frameworks for handling uncertain covariates. Neural network modeling, by contrast, arose in domains like computer vision \citep{Simonyan2014, Szegedy2014, He2016} and natural language processing \citep{Devlin2019, Brown2020, OpenAI2023, Touvron2023} where input noise is negligible, and its standard architectures have little mechanism for directly incorporating input uncertainties. When applied to astronomical data, this simplification can lead to systematic errors that are difficult to diagnose \citep{Kelly2007, Ting2025_PaperI}.

Astronomy operates in a regime that differs from most machine learning applications. Astronomical observations often have signal-to-noise ratios (SNR) of order 10, yet we seek inference precision at the percent level or better. This tension between noisy inputs and stringent precision requirements creates an unique challenge for neural network inference.

This phenomenon is known as attenuation bias \citep{Spearman1904, Frost2002, Kelly2007}. In Paper~I of this series \citep{Ting2025_PaperI}, we demonstrated this effect in detail for linear regression and explored its implications for astronomical data analysis. Attenuation bias manifests as a compression of the predicted dynamic range: high values are consistently predicted too low, while low values are predicted too high. The effect depends on the ratio of input measurement error to intrinsic signal range, and it is independent of sample size. As a rule of thumb, attenuation bias becomes important at input SNR of order 10, especially at low-dimensional problems. The exact magnitude depends on the dimensionality and correlation structure of the inputs, but the bias is never zero when input uncertainties are present. We refer interested readers to Paper~I for detailed derivations and further discussion of attenuation bias in the context of linear regression.

The consequences of attenuation bias are not merely increased scatter in predictions. When we train a regression model to predict some quantity from noisy inputs, treating the observed inputs as exact is a reasonable approximation only when measurement errors are small compared to the intrinsic variation in the data. When measurement errors constitute even a modest fraction of the signal, a systematic compression of the predicted dynamic range emerges---high values are underestimated and low values are overestimated---a bias that cannot be removed by collecting more data or by using more sophisticated models. It is a fundamental consequence of treating noisy measurements as if they were exact.

For linear models, the statistics literature has long recognized this problem and developed solutions such as Deming regression \citep{Deming1943, Fuller1987, Carroll2006, Ting2025b}. The key insight is to treat the true input values as latent variables to be estimated alongside model parameters \citep{Loredo2004, Kelly2007}. Rather than taking the noisy observations at face value, the model simultaneously infers what the true inputs likely were and learns the relationship between those true inputs and the outputs. The present paper extends this idea to neural networks. We show that neural networks suffer from the same attenuation bias as linear regression, and that the latent variable approach provides a general solution applicable to neural networks.

We introduce LatentNN, a practical implementation using modern automatic differentiation frameworks, and validate it on one-dimensional regression, multivariate inputs, and spectral applications. Section~\ref{sec:review} briefly reviews attenuation bias following Paper~I notation. Section~\ref{sec:nn_attenuation} demonstrates that neural networks exhibit the same bias. Section~\ref{sec:latentnn} presents the LatentNN formulation and results, first for the one-dimensional case, then extending to multivariate inputs. Section~\ref{sec:spectra} applies the method to stellar spectra. Section~\ref{sec:discussion} discusses limitations, extensions, and connections to hierarchical Bayesian modeling, followed by conclusions in Section~\ref{sec:conclusions}.

\begin{figure}[t]
    \centering
    \includegraphics[width=\columnwidth]{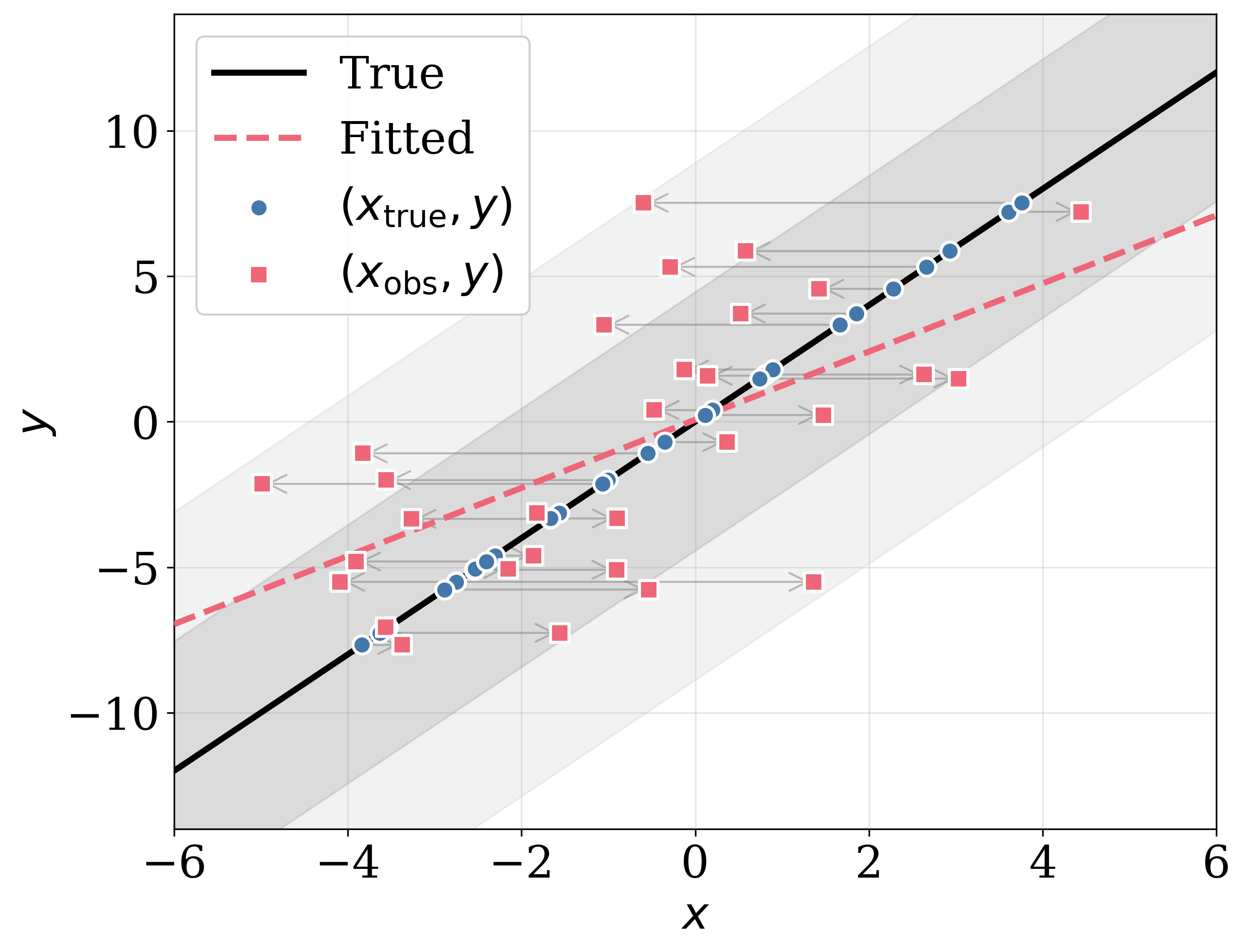}
    \caption{Schematic illustration of attenuation bias. Blue circles show true positions $(x_{\rm true}, y)$ following $y = 2x$ (black line); red squares show observed positions $(x_{\rm obs}, y)$ after adding measurement error with $\mathrm{SNR}_x = 1$. Grey arrows indicate horizontal displacement from true to observed $x$ positions. The tilted grey shaded bands show the $\pm 1\sigma_x$ (darker) and $\pm 2\sigma_x$ (lighter) uncertainty regions around the true relationship. A linear fit to the observed data (red dashed line) yields a shallower slope, demonstrating how input measurement errors systematically bias regression coefficients toward zero.}
    \label{fig:schematic}
\end{figure}

\section{Attenuation Bias}
\label{sec:review}

To understand attenuation bias, consider a simple regression problem where we want to learn the relationship between an input variable $x$ and an output variable $y$. Suppose the true relationship is $y_{\rm true} = f(x_{\rm true})$, which could be linear or nonlinear. In practice, we never observe the true values directly. Instead, we measure $x_{\rm obs} = x_{\rm true} + \delta_x$, where $\delta_x$ represents the measurement error. If these errors are unbiased with variance $\sigma_x^2$, then $\delta_x \sim \mathcal{N}(0, \sigma_x^2)$. The naive approach in machine learning is to train on the observed pairs $(x_{\rm obs}, y_{\rm obs})$, treating $x_{\rm obs}$ as if it were the true value. This seemingly innocuous simplification leads to systematic bias.

Measurement errors in $x$ artificially inflate the observed variance. If the true values have variance $\sigma_{\rm range}^2 = {\rm Var}(x_{\rm true})$, representing the intrinsic spread or dynamic range of the data, then the observed variance is
\begin{equation}
{\rm Var}(x_{\rm obs}) = \sigma_{\rm range}^2 + \sigma_x^2.
\end{equation}
This stretches the data horizontally without a corresponding vertical stretch, because the measurement errors in $x$ do not affect $y$. The observed distribution appears wider in the $x$-direction than the true distribution, while the $y$-extent remains unchanged. Any regression method treating $x_{\rm obs}$ as exact will therefore fit a slope that is systematically shallower than the true relationship. This weakening of the apparent relationship is why attenuation bias is also known as regression dilution.

This effect can be visualized by plotting data points and their regression line. Figure~\ref{fig:schematic} illustrates this with a simple example: true data points (blue) following a linear relationship $y = 2x$ are scattered horizontally by measurement error to their observed positions (red), with arrows indicating the displacement. Grey tilted shaded bands show the $\pm 1\sigma_x$ and $\pm 2\sigma_x$ uncertainty regions around the true relationship. A regression line fit to the observed data (red dashed) has a shallower slope than the true underlying relationship.

For linear regression with true relationship $y = \beta x$, where $\beta$ is the true slope, we can quantify this effect precisely (see Appendix~\ref{app:attenuation} for the full derivation). The ordinary least squares estimator $\hat{\beta}$ has expected value
\begin{equation}
\mathbb{E}[\hat{\beta}] = \beta \cdot \lambda_\beta, \quad \text{where} \quad \lambda_\beta = \frac{1}{1 + (\sigma_x/\sigma_{\rm range})^2}.
\label{eq:attenuation}
\end{equation}
The quantity $\lambda_\beta$ is called the attenuation factor. It depends only on the ratio of measurement error to intrinsic signal range, $\sigma_x/\sigma_{\rm range}$. When measurement errors are small compared to the signal range, $\lambda_\beta$ is close to unity and the bias is negligible. When measurement errors are comparable to the signal range, $\lambda_\beta$ can be much less than unity, indicating large bias. Since $\lambda_\beta < 1$, predictions are systematically pulled toward the mean: high values are predicted too low, while low values are predicted too high. In the extreme case where measurement errors dominate ($\sigma_x \gg \sigma_{\rm range}$), $\lambda_\beta$ approaches zero and the model predicts the mean for all inputs, obscuring the true relationship.

Consider some representative cases in the one-dimensional setting:
\begin{itemize}
\item At $\sigma_x/\sigma_{\rm range} = 0.1$ (equivalently, SNR $= \sigma_{\rm range}/\sigma_x = 10$), the attenuation factor is $\lambda_\beta \approx 0.99$, corresponding to 1\% bias.
\item At $\sigma_x/\sigma_{\rm range} = 0.33$ (SNR $\approx 3$), the attenuation factor is $\lambda_\beta \approx 0.90$, corresponding to 10\% bias.
\item At $\sigma_x/\sigma_{\rm range} = 1.0$ (SNR $= 1$), the attenuation factor is $\lambda_\beta = 0.50$, corresponding to 50\% bias.
\end{itemize}
At SNR of 10 or below, a regime common in astronomical observations, the bias reaches percent level or more.

Several properties of this bias are worth emphasizing. First, the bias depends only on the ratio $\sigma_x/\sigma_{\rm range}$, not on the absolute values of either quantity. Second, it is independent of sample size $n$ and cannot be resolved by collecting more data. Even with infinite training samples, the bias remains. Third, the magnitude of attenuation bias depends only on errors in $x$, not errors in $y$. Errors in $y$ add scatter to the predictions but do not systematically bias them, because positive and negative errors average out over the dataset. In other words, obtaining better or more accurate labels does not solve the problem.\footnote{This study focuses on discriminative models that map observables to labels. When labels are more accurate than inputs, generative models that predict observables from labels can mitigate attenuation bias, as the roles of $x$ and $y$ are reversed. See Paper~I for further discussion.}

This asymmetry between $x$ and $y$ arises because they play fundamentally different roles in regression: we predict $y$ given $x$, not the reverse. Errors in $x$ change the question being asked---we are predicting from a corrupted version of the input---which systematically weakens the apparent relationship. However, as we will see in Section~\ref{sec:latentnn}, while $\sy$ does not affect the bias itself, it does play a role in correcting the bias: the Deming regression solution \citep{Deming1943, GolubVanLoan1980, Ting2025b} and its neural network generalization require specifying both $\sx$ and $\sy$ to properly weight the loss terms.

It is worth noting that this bias is sometimes mistakenly attributed to insufficient training data at the extremes or to noisy labels. Observing that models underpredict extreme values, one might assume the problem lies in having too few training examples at the extremes or in noisy labels. However, attenuation bias has nothing to do with either of these issues. The bias arises purely from measurement errors in the input, and it persists even with perfectly balanced training data and perfect labels.

\begin{figure*}[t]
    \centering
    \includegraphics[width=1.8\columnwidth]{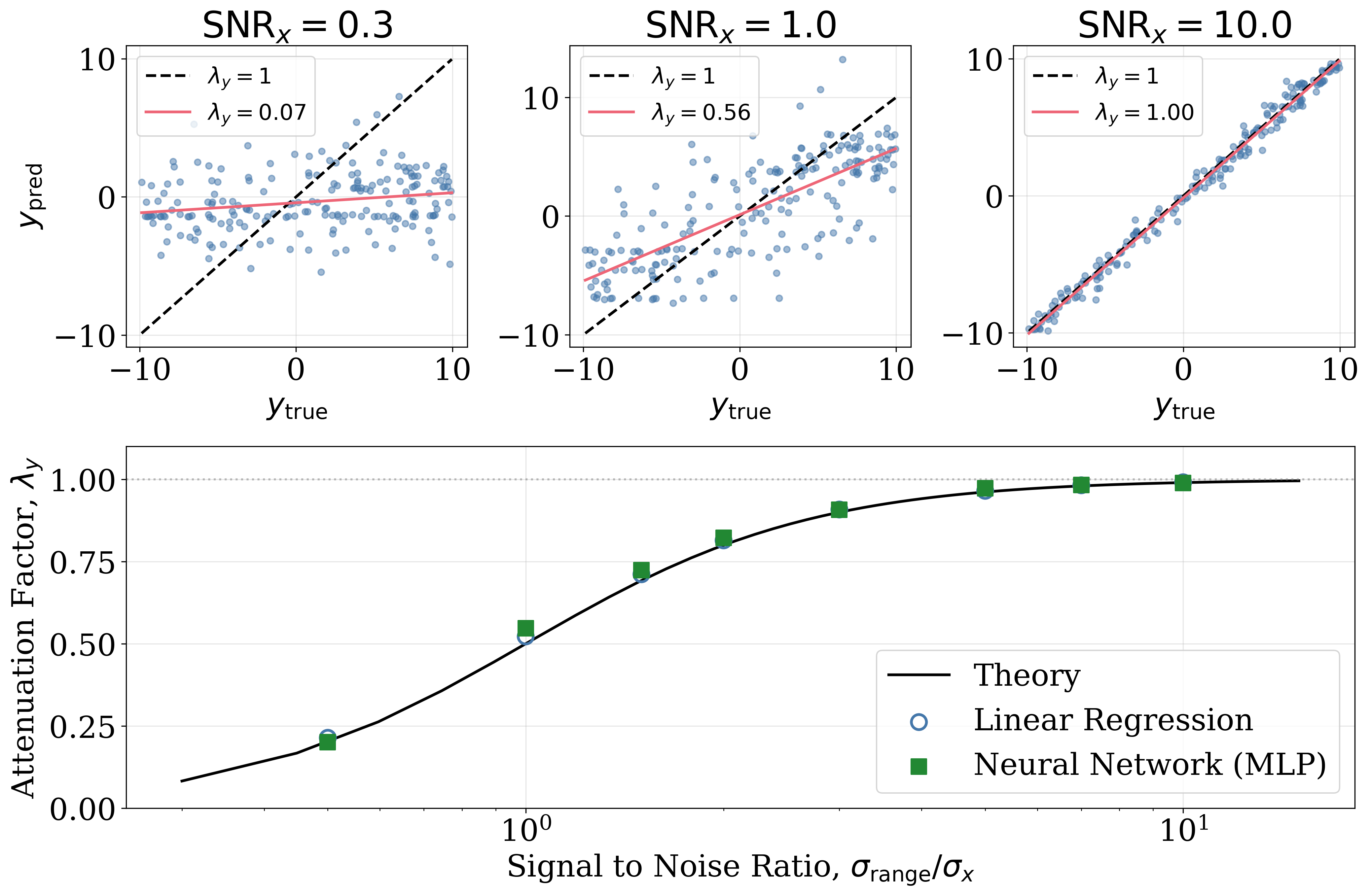}
    \caption{Attenuation bias in neural networks for the true function $y = 2x$. \textbf{Top:} Predicted versus true $y$ on held-out test data for an MLP (2 hidden layers, 64 units) at three $\mathrm{SNR}_x$ values. At $\mathrm{SNR}_x = 0.3$, predictions collapse to $\lambda_y \approx 0.1$; at $\mathrm{SNR}_x = 1$, $\lambda_y \approx 0.5$; at $\mathrm{SNR}_x = 3$, $\lambda_y \approx 0.9$. \textbf{Bottom:} Attenuation factor $\lambda_y$ versus $\mathrm{SNR}_x$ for linear regression (blue open circles) and neural network (green filled squares). Both follow the theoretical curve $\lambda_y = 1/(1 + \mathrm{SNR}_x^{-2})$ (black solid line), demonstrating that model complexity does not protect against attenuation bias.}
    \label{fig:mlp_attenuation}
    \end{figure*}

\section{Attenuation Bias in Neural Networks}
\label{sec:nn_attenuation}

The previous section reviewed attenuation bias in the context of linear regression, where the effect can be derived analytically. Paper I further argued that the same bias extends to polynomial and nonlinear models. However, for nonlinear models or high-dimensional inputs, the attenuation factor can no longer be expressed as a simple function of $\sigma_x$ and $\sigma_{\rm range}$. The relationship between measurement error and attenuation becomes more complex, depending on the specific functional form.

This complexity makes it difficult to predict and therefore analytically correct the bias in closed form for general machine learning models. Here we first demonstrate empirically that multi-layer perceptrons (MLPs) exhibit the same qualitative behavior. MLPs, also known as fully connected neural networks \citep{Cybenko1989, HornikStinchcombeWhite1989}, are the simplest neural network architecture: each neuron in one layer is connected to every neuron in the next layer, with nonlinear activation functions between layers. Despite their simplicity, MLPs remain widely used in astronomy \citep[e.g.,][]{Odewahn1992, Storrie-Lombardi1992, Collister2004, Guo2020, Wong2020, Wang2021, Li2022_LAMOST, Vynatheya2022, Winecki2024} and serve as the foundation for more complex architectures.

To test whether neural networks suffer from attenuation bias, we conduct numerical experiments following the setup of Paper~I. We consider the true function $y = 2x$, chosen for direct comparison with the linear theory, and sample $N_{\rm train} = 1000$ training values and $N_{\rm test} = 200$ test values of $\xtrue$ uniformly from $[-5, 5]$. We add Gaussian noise to obtain $\xobs = \xtrue + \delta_x$ where $\delta_x \sim \mathcal{N}(0, \sigma_x^2)$, and vary $\sx$ to test different SNR regimes. We define $\mathrm{SNR}_x \equiv \srange/\sx$ as the signal-to-noise ratio in the input.

We train two models on each dataset. The first is ordinary least squares linear regression. The second is an MLP with 2 hidden layers of 64 units each and ReLU activations, trained for 20000 epochs using the Adam optimizer with learning rate 0.03.

For neural networks, we cannot directly measure a single coefficient $\beta$ as we did for linear regression in Section~\ref{sec:review}. The MLP used here has over 4000 parameters distributed across multiple layers, rather than a single interpretable slope. Instead, we quantify attenuation through the predictions themselves. Here, predictions are always made using the observed (noisy) inputs: $y_{\rm pred} = f_{\boldsymbol{\theta}}(x_{\rm obs})$, not the unknown true inputs. We define $\lambda_y$ as the slope of the regression of $y_{\rm pred}$ against $y_{\rm true}$. If a model suffers from attenuation bias, its predictions will be compressed toward the mean: when we plot $y_{\rm pred}$ versus $y_{\rm true}$, the slope will be less than unity. A value of $\lambda_y = 1$ indicates no attenuation, while $\lambda_y < 1$ indicates that predictions are systematically pulled toward the sample mean.

As discussed in Paper~I, for linear regression in one dimension with true relationship $y = \beta x$, these two measures of attenuation are equivalent. The predicted value is $\hat{y} = \hat{\beta} x_{\rm obs}$, and when we regress predictions against true values, the attenuation in the slope estimate $\hat{\beta}$ translates directly to attenuation in the predictions. In this case, $\lambda_y = \lambda_\beta = 1/(1 + (\sigma_x/\sigma_{\rm range})^2)$. For nonlinear models or multivariate inputs, $\lambda_y$ remains a useful empirical measure of how much the model compresses the dynamic range of predictions.

Figure~\ref{fig:mlp_attenuation} shows the results. The top row shows MLP predictions on held-out test data at three specific $\mathrm{SNR}_x$ values: at $\mathrm{SNR}_x = 0.3$ (severe noise), predictions are compressed to $\lambda_y \approx 0.1$; at $\mathrm{SNR}_x = 1$ (noise comparable to signal range), $\lambda_y \approx 0.5$; and at $\mathrm{SNR}_x = 3$, $\lambda_y \approx 0.9$. The bottom panel shows that both linear regression and the MLP follow the theoretical attenuation curve $\lambda_y = 1/(1 + (\sx/\srange)^2)$ across the full range of $\mathrm{SNR}_x$. For the linear true function $y = 2x$, the MLP effectively learns a linear mapping, so it follows the same theoretical curve as linear regression. The important point is that our MLP here, despite having thousands of parameters and the flexibility to fit arbitrary nonlinear functions, exhibits the same systematic bias as the much simpler linear regression. As we argued in Paper~I, model complexity does not protect against attenuation bias.

Regularization does not resolve the issue either. One might hope that $L_2$ regularization (weight decay), which penalizes large weights, could counteract the attenuation \citep{Tikhonov1963, HoerlKennard1970, Bishop2006, Ting2025b}. However, regularization addresses overfitting by constraining model complexity, not systematic bias in the inputs. The bias arises from the mismatch between $\xobs$ and $\xtrue$, which regularization cannot correct. We verified this empirically by sweeping over a range of $L_2$ regularization strengths: the attenuation factor $\lambda_y$ remains unchanged regardless of the regularization parameter.

\section{LatentNN: Correcting Attenuation Bias with Latent Variables}
\label{sec:latentnn}

Having established that neural networks suffer from attenuation bias, we now develop a method to correct it. The approach builds on a classical idea from the statistics literature on \emph{errors-in-variables} regression \citep{Carroll1995, Carroll2006, Bishop2006}: rather than treating noisy observations as exact, we treat the unknown true values as latent variables to be estimated alongside the model parameters. For linear regression, this leads to Deming regression \citep{Deming1943}. Here we show that the same principle extends naturally to neural networks, yielding a practical correction method we call LatentNN.

\subsection{The Latent Variable Solution in Linear Regression}

Standard regression treats $x_{\rm obs}$ as exact, which leads to bias. An alternative approach recognizes that $x_{\rm true}$ is an unknown quantity for each data point. Rather than treating the observed values as truth, we can treat the true values as latent variables to be estimated alongside the model parameters.

Consider observation $i$. We assume that the observed input $x_{{\rm obs},i}$ is a noisy measurement of the true value $x_{{\rm true},i}$, and the observed output $y_{{\rm obs},i}$ is a noisy measurement of the true output $y_{{\rm true},i}$, which follows the linear relationship $y_{{\rm true},i} = \beta x_{{\rm true},i}$. Mathematically,
\begin{align}
x_{{\rm obs},i} &= x_{{\rm true},i} + \delta_{x,i}, \quad \delta_{x,i} \sim \mathcal{N}(0, \sigma_x^2), \\
y_{{\rm obs},i} &= \beta x_{{\rm true},i} + \delta_{y,i}, \quad \delta_{y,i} \sim \mathcal{N}(0, \sigma_y^2).
\end{align}
Here $\sigma_y^2$ is the known measurement uncertainty in the output. In principle, one could add a jitter term to account for intrinsic scatter about the true relationship, but we assume pure measurement noise for simplicity.

Given these assumptions, the joint probability of observing both $x_{{\rm obs},i}$ and $y_{{\rm obs},i}$ given the true value $x_{{\rm true},i}$ and the model parameter $\beta$ is
\begin{align}
p(x_{{\rm obs},i}, y_{{\rm obs},i} | x_{{\rm true},i}, \beta) &= \mathcal{N}(x_{{\rm obs},i} | x_{{\rm true},i}, \sigma_x^2) \nonumber \\
&\quad \times \mathcal{N}(y_{{\rm obs},i} | \beta x_{{\rm true},i}, \sigma_y^2).
\label{eq:joint_likelihood}
\end{align}
Here $\mathcal{N}(x | \mu, \sigma^2)$ denotes the probability density of observing $x$ from a Gaussian distribution with mean $\mu$ and variance $\sigma^2$. The first factor is the likelihood of the observed input given the true input. The second factor is the likelihood of the observed output given the model prediction. Conditioning on $x_{{\rm true},i}$ makes $x_{{\rm obs},i}$ and $y_{{\rm obs},i}$ independent because they scatter due to independent noise sources.

Over all $N$ observations, the full likelihood is
\begin{align}
\mathcal{L}(\beta, \{x_{{\rm true},i}\}) = \prod_{i=1}^{N} &\mathcal{N}(x_{{\rm obs},i} | x_{{\rm true},i}, \sigma_x^2) \nonumber \\
&\times \mathcal{N}(y_{{\rm obs},i} | \beta x_{{\rm true},i}, \sigma_y^2).
\end{align}

This formulation introduces $N+1$ parameters---the slope $\beta$ plus the true value $x_{{\rm true},i}$ for each of the $N$ data points---for $2N$ observed quantities. 

At first glance this appears underdetermined. However, the problem is well-posed for two reasons. First, the linear relationship provides an inductive bias that constrains all $x_{{\rm true},i}$ to be consistent with a single slope. The true values are not independent; they must all be consistent with the same underlying linear relationship. Second, the term involving $x_{\rm obs}$ acts as a regularizer, preventing the $x_{\rm true}$ values from drifting arbitrarily far from the observations. This balance between fitting the model and staying close to the data leads to the Deming regression solution, which has a closed form (see Appendix~\ref{app:deming}).

This approach of treating unknown true values as parameters to be inferred is an example of hierarchical modeling \citep{Loredo2004, Kelly2007, Gelman2013}. In this hierarchical structure, the model parameter $\beta$ sits at the top level, the true values $x_{{\rm true},i}$ form the intermediate level, and the actual measurements $(x_{\rm obs}, y_{\rm obs})$ sit at the bottom level. The model constrains the intermediate level, while the intermediate level generates the observations.

The resulting Deming regression estimator (derived in Appendix~\ref{app:deming}) recovers the true slope $\beta$ without attenuation. When the error variances $\sigma_x^2$ and $\sigma_y^2$ are known, defining the error variance ratio $r = \sigma_y^2/\sigma_x^2$, the estimator is
\begin{equation}
\hat{\beta}_{\rm Deming} = \frac{S_{yy} - r S_{xx} + \sqrt{(S_{yy} - r S_{xx})^2 + 4 r S_{xy}^2}}{2 S_{xy}},
\end{equation}
where $S_{xx} = \sum_i (x_{{\rm obs},i} - \bar{x})^2$, $S_{yy} = \sum_i (y_{{\rm obs},i} - \bar{y})^2$, and $S_{xy} = \sum_i (x_{{\rm obs},i} - \bar{x})(y_{{\rm obs},i} - \bar{y})$ are the sum of squared deviations and cross-products, with $\bar{x}$ and $\bar{y}$ denoting the sample means. In the limit $\sigma_x \to 0$ (equivalently $r \to \infty$), this reduces to the ordinary least squares estimator $\hat{\beta}_{\rm OLS} = S_{xy}/S_{xx}$. When $\sigma_x > 0$, the Deming estimator yields a steeper slope than OLS, correcting for the attenuation \citep[see also][]{Ting2025_PaperI}.

The solution for the optimal $x_{{\rm true},i}$ provides further intuition. For fixed $\beta$, the optimal estimate is a weighted average
\begin{equation}
x_{{\rm true},i} = w_x \cdot x_{{\rm obs},i} + w_y \cdot \frac{y_{{\rm obs},i}}{\beta},
\end{equation}
where the weights are (see Appendix~\ref{app:deming} for derivation)
\begin{equation}
w_x = \frac{r}{r + \beta^2}, \quad w_y = \frac{\beta^2}{r + \beta^2}.
\end{equation}
The weights sum to unity. When $\sigma_x$ is small compared to $\sigma_y/\beta$ (i.e., $r$ is large), the weight $w_x$ dominates and we trust the direct $x$ measurement. When $\sigma_x$ is large (i.e., $r$ is small), the weight $w_y$ dominates and we infer the true $x$ from the $y$ measurement and the model relationship. This optimal weighting is what allows the method to correct for attenuation bias. By appropriately balancing direct measurements against the constraints imposed by the model, we recover estimates of both the true parameter values and the true underlying data points.

\subsection{The LatentNN Formulation}

The linear form of Deming regression was not essential---what mattered was the hierarchical structure where latent values generate observations and the model constrains the latent values. The linear form permits an analytic solution, but the underlying principle of joint optimization over model parameters and latent true values extends directly to neural networks.

We replace the linear model $\beta x$ with a neural network $f_{\boldsymbol{\theta}}(x)$, where $\boldsymbol{\theta}$ represents all the network weights and biases. The hierarchical structure remains the same: the network parameters $\boldsymbol{\theta}$ sit at the top level (analogous to $\beta$), the latent values $x_{{\rm latent},i}$ form the intermediate level (analogous to $x_{{\rm true},i}$), and the observations $(x_{\rm obs}, y_{\rm obs})$ sit at the bottom level. The likelihood has exactly the same form as in the linear case,
\begin{align}
\mathcal{L}(\boldsymbol{\theta}, \{x_{{\rm latent},i}\}) = \prod_{i=1}^{N} &\mathcal{N}(x_{{\rm obs},i} | x_{{\rm latent},i}, \sigma_{x,i}^2) \nonumber \\
&\times \mathcal{N}(y_{{\rm obs},i} | f_{\boldsymbol{\theta}}(x_{{\rm latent},i}), \sigma_{y,i}^2),
\end{align}
with $f_{\boldsymbol{\theta}}(x_{{\rm latent},i})$ replacing $\beta x_{{\rm true},i}$. Here $\sigma_{x,i}$ and $\sigma_{y,i}$ denote the per-sample uncertainties, which may vary across observations.

Taking the negative log-likelihood, rearranging, and absorbing constants, we obtain the loss function
\begin{align}
\mathcal{J}(\boldsymbol{\theta}, \{x_{{\rm latent},i}\}) &= \sum_{i=1}^{N} \frac{(y_{{\rm obs},i} - f_{\boldsymbol{\theta}}(x_{{\rm latent},i}))^2}{\sigma_{y,i}^2} \nonumber \\
&\quad + \sum_{i=1}^{N} \frac{(x_{{\rm obs},i} - x_{{\rm latent},i})^2}{\sigma_{x,i}^2}.
\label{eq:latentnn_loss}
\end{align}
Comparing to the Deming regression objective in Appendix~\ref{app:deming}, we see the same structure: a prediction loss term weighted by $1/\sigma_{y,i}^2$ and a latent likelihood term weighted by $1/\sigma_{x,i}^2$. The balance between these terms is controlled by the ratio of uncertainties, which determines how much the latent values can deviate from the observations to improve predictions. This is the same trade-off we saw in the weighted average formula for $x_{{\rm true},i}$: trust the direct measurement when it is precise, but infer the true value from the model when the measurement is noisy. The uncertainties $\sigma_{x,i}$ and $\sigma_{y,i}$ must be specified for proper correction. In the experiments that follow, to facilitate our study and visualization of the results, we simplify to homogeneous uncertainties $\sigma_x$ and $\sigma_y$ across all samples. 

The optimization problem is to find
\begin{equation}
\hat{\boldsymbol{\theta}}, \{\hat{x}_{{\rm latent},i}\} = \underset{\boldsymbol{\theta}, \{x_{{\rm latent},i}\}}{\arg\min} \; \mathcal{J}(\boldsymbol{\theta}, \{x_{{\rm latent},i}\}).
\label{eq:latentnn_argmin}
\end{equation}
For a dataset with $N$ samples and $p$ input dimensions, this involves optimizing over the network parameters $\boldsymbol{\theta}$ plus $N \times p$ latent variables. The total number of parameters is thus much larger than in standard neural network training, scaling with the training set size.

The key difference from the linear case is that Equation~\ref{eq:latentnn_argmin} has no closed-form solution. In Deming regression, the linear structure allows us to solve for both $\hat{\beta}$ and $\{\hat{x}_{{\rm true},i}\}$ analytically (see Appendix~\ref{app:deming}). For a neural network $f_{\boldsymbol{\theta}}$, the nonlinear dependence on $\boldsymbol{\theta}$ precludes an analytic solution, and we must resort to numerical optimization.

\begin{figure*}[t]
    \centering
    \includegraphics[width=2\columnwidth]{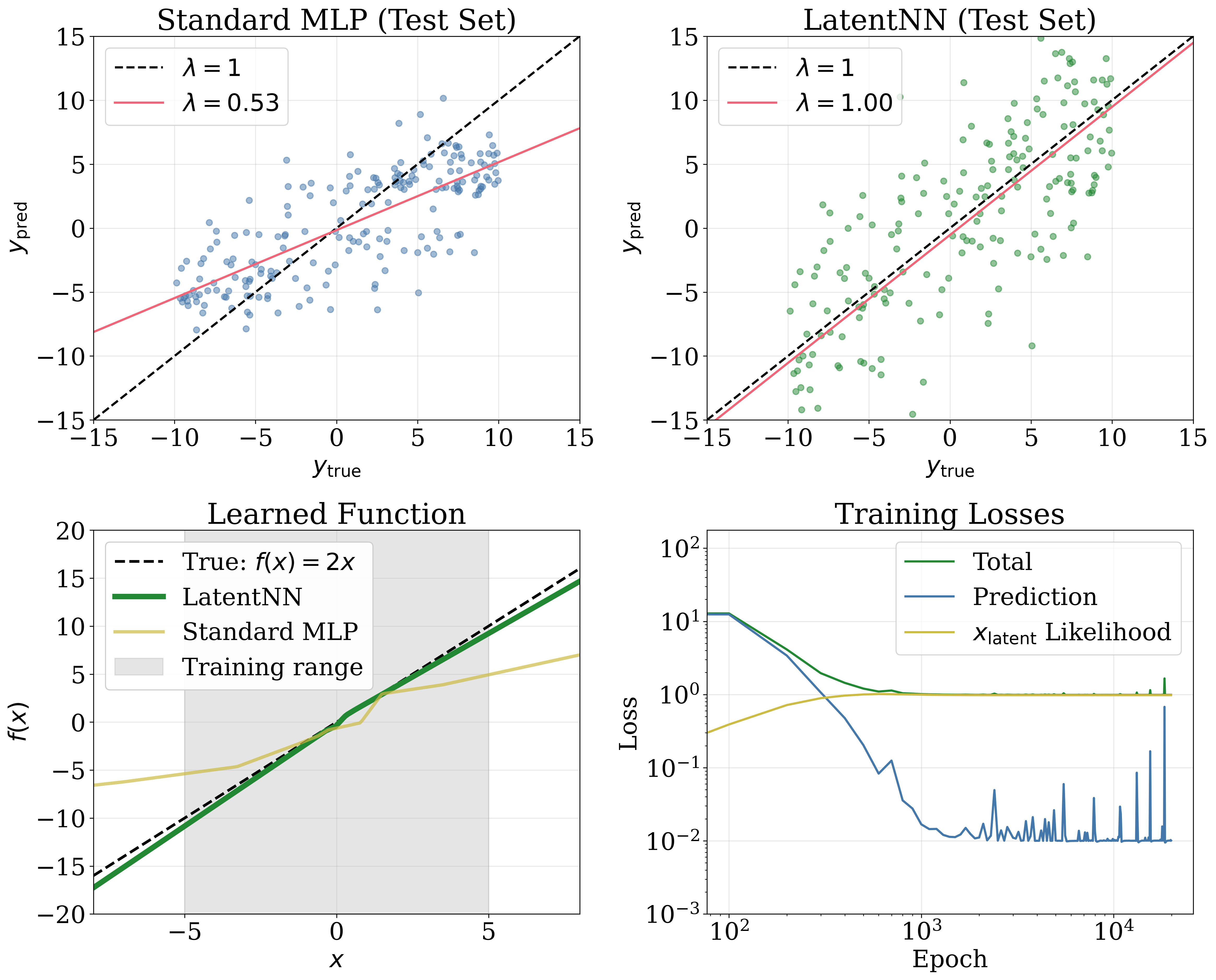}
    \caption{LatentNN demonstration at $\mathrm{SNR}_x=1$. \textbf{Top:} Predicted versus true $y$ on the test set for standard neural networks with the MLP arhitecture (left, $\lambda_y \approx 0.5$) and LatentNN (right, $\lambda_y \approx 1$). LatentNN corrects the attenuation bias. \textbf{Bottom left:} Learned functions. The standard MLP (orange) learns an attenuated slope, while LatentNN (green) recovers the true function $f(x) = 2x$ (black dashed). The gray shaded region indicates the noiseless training x-value range. \textbf{Bottom right:} LatentNN training losses. The prediction loss decreases while the $x_{\rm latent}$ likelihood increases, reflecting the trade-off as latent values shift from noisy observations toward true values.}
    \label{fig:latentnn_1d}
\end{figure*}

\begin{figure*}[t]
    \centering
    \includegraphics[width=\textwidth]{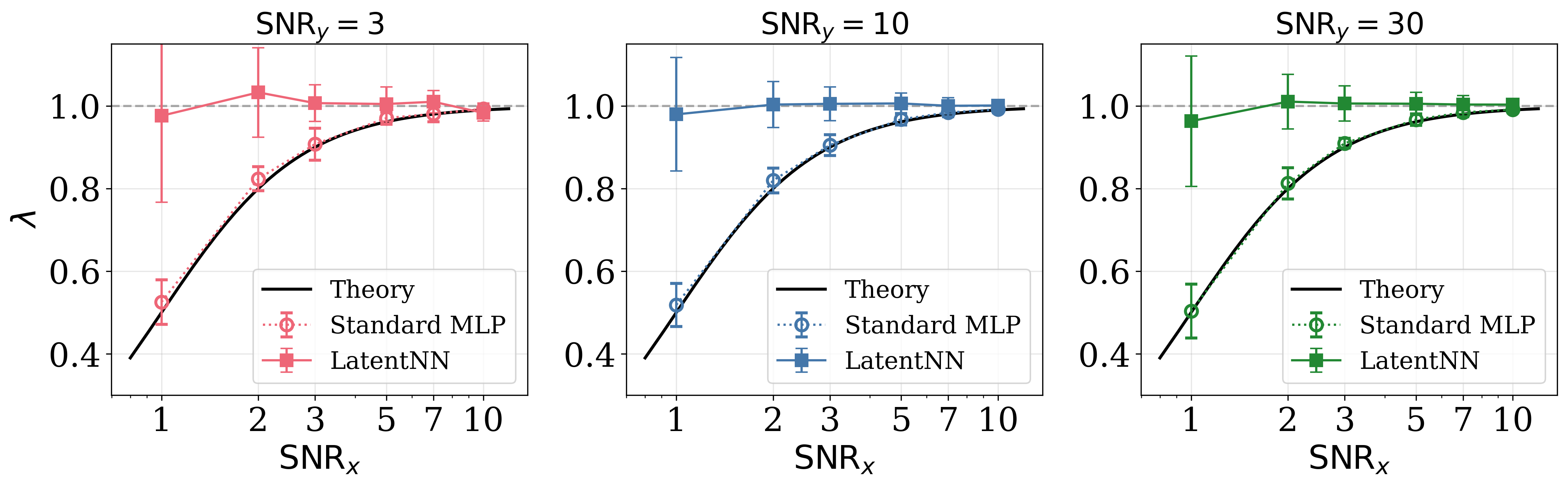}
    \caption{Attenuation factor $\lambda_y$ versus $\mathrm{SNR}_x$ for three $\mathrm{SNR}_y$ levels (3, 10, 30; different colors). Error bars show standard deviation over 8 runs with different random seeds, using a three-fold data split (train/validation/test) to avoid data leakage in hyperparameter selection. Standard MLP (open circles, dotted) follows the theoretical attenuation curve (black solid) regardless of $\mathrm{SNR}_y$. LatentNN (filled squares, solid) maintains mean $\lambda_y \approx 1$ across all tested SNR values. Even at $\mathrm{SNR}_x = 1$, LatentNN achieves unbiased predictions on average, though with increased run-to-run scatter at lower $\mathrm{SNR}_x$ as the problem becomes increasingly ill-conditioned.}
    \label{fig:attenuation_comparison}
\end{figure*}

\subsection{Implementation}

We solve Equation~\ref{eq:latentnn_argmin} using gradient descent, the standard optimization algorithm for neural networks. Let $\Phi = (\boldsymbol{\theta}, \{x_{{\rm latent},i}\})$ denote the full set of parameters to be optimized. At each iteration $t$, we update
\begin{equation}
\Phi^{(t+1)} = \Phi^{(t)} - \eta \nabla_\Phi \mathcal{J},
\end{equation}
where $\eta$ is the learning rate and $\nabla_\Phi \mathcal{J}$ is the gradient of the loss with respect to all parameters. This is the same update rule used in standard neural network training, but with $\Phi$ now including both the network weights $\boldsymbol{\theta}$ and the latent inputs $\{x_{{\rm latent},i}\}$. This represents a conceptual shift from typical neural network applications, where we optimize only over $\boldsymbol{\theta}$. Here, we treat the observed inputs as noisy measurements rather than ground truth, and optimize jointly over $\boldsymbol{\theta}$ and the latent inputs.

In modern deep learning frameworks such as PyTorch or JAX, this is straightforward to implement. The latent values $x_{{\rm latent},i}$ are declared as additional learnable parameters, and the framework automatically computes gradients via backpropagation \citep{RumelhartHintonWilliams1986, Lecun1998}.

Several practical considerations affect performance. First, initialization matters. We initialize the latent values at the observed values:
\begin{equation}
x_{{\rm latent},i}^{(0)} = x_{{\rm obs},i}.
\label{eq:initialization}
\end{equation}
In practice, this means declaring a learnable parameter array of shape $(N, p)$ initialized to the observed input matrix. This ensures that the initial loss is reasonable and provides a starting point consistent with our probabilistic model.

Second, additional $L_2$ regularization (weight decay) on the network parameters $\boldsymbol{\theta}$ is important. The loss function in Equation~\ref{eq:latentnn_loss} has two terms: a prediction loss that encourages the network to fit the data, and a latent likelihood term that keeps the latent values close to the observations. With $N \times p$ latent variables plus thousands of network parameters, the optimization can find solutions where the network overfits to specific latent value configurations rather than learning a generalizable mapping. Weight decay on the network parameters prevents this by penalizing large weights, encouraging smoother functions that generalize better. Weight decay is applied only to $\boldsymbol{\theta}$, not to the latent values $x_{\rm latent}$---the latent values are already constrained by the latent likelihood term.

\section{Numerical Experiments}
\label{sec:results}

We now validate the LatentNN approach on test cases where the true relationship and measurement errors are known, allowing direct comparison between estimated and true attenuation factors. We begin with the one-dimensional case then extend to multivariate inputs with correlated features.

\subsection{One-Dimensional Case}

We use the same data setup and network architecture as in Section~\ref{sec:nn_attenuation}: the true relationship $y = 2x_{\rm true}$ with 1000 training samples, and an MLP with 2 hidden layers of 64 units trained for 20000 epochs. For LatentNN, we add a learnable parameter $x_{{\rm latent},i}$ for each training sample, initialized to $x_{{\rm obs},i}$. 

To ensure fair comparison with no data leakage, we use a three-fold data split: 1000 samples for training, 200 for validation (hyperparameter tuning), and 200 for final evaluation. The validation set is used exclusively for selecting hyperparameters---for LatentNN we maximize $\lambda_y$ on the validation set, while for the standard MLP we minimize validation loss. Final performance metrics are computed on the held-out test set, which is never used during hyperparameter selection and the training process. 

Figure~\ref{fig:latentnn_1d} shows a representative comparison at $\mathrm{SNR}_x = 1$, using $\mathrm{SNR}_y = 10$ as in Section~\ref{sec:nn_attenuation}. At $\mathrm{SNR}_x = 1$, the measurement error equals the signal range ($\sigma_x = \sigma_{\rm range}$)---an extreme regime where the noise is 100\% of the data's dynamic range. The top panels show predicted versus true values on the held-out test set. The standard MLP shows $\lambda_y \approx 0.5$, demonstrating large attenuation bias. LatentNN recovers $\lambda_y \approx 1$, correcting the bias. The bottom-left panel shows the learned functions: the standard MLP learns an attenuated slope, while LatentNN recovers the true relationship $f(x) = 2x$.

The bottom-right panel shows the training dynamics for LatentNN. At initialization, $x_{\rm latent} = x_{\rm obs}$, so the $x_{\rm latent}$ likelihood term starts near zero. As training proceeds, the prediction loss decreases as the network learns the mapping, but the $x_{\rm latent}$ likelihood increases. This increase is expected and desirable: to achieve accurate predictions, the latent values must shift from the noisy observations $x_{\rm obs}$ toward the true values $x_{\rm true}$, which increases the likelihood penalty. The total loss decreases overall because the improvement in prediction accuracy outweighs the likelihood cost. At convergence, the optimization has found the balance point where the latent values have moved sufficiently toward their true positions to enable unbiased prediction.

To systematically evaluate performance, we repeat each experiment 8 times with different random seeds, using the same three-fold data split (train/validation/test) for each run. Figure~\ref{fig:attenuation_comparison} shows $\lambda_y$ versus $\mathrm{SNR}_x$ across a range of noise levels from $\mathrm{SNR}_x = 1$ to $\mathrm{SNR}_x = 10$, testing three $\mathrm{SNR}_y$ values (3, 10, and 30). As expected from the theory in Section~\ref{sec:review}, varying $\mathrm{SNR}_y$ has little effect on attenuation---the bias depends primarily on input noise. The standard MLP follows the theoretical attenuation curve $\lambda_y = 1/(1 + (\sigma_x/\sigma_{\rm range})^2)$ regardless of $\mathrm{SNR}_y$. LatentNN maintains $\lambda_y \approx 1$ across all tested $\mathrm{SNR}_x$ and $\mathrm{SNR}_y$ values. LatentNN provides robust correction even at $\mathrm{SNR}_x = 1$, though with increased run-to-run scatter at lower $\mathrm{SNR}_x$ values as the problem becomes increasingly ill-conditioned.

The mechanism of correction differs between SNR regimes. At high SNR where $\sigma_x/\sigma_{\rm range} \ll 1$, the $x_{\rm latent}$ likelihood term in Equation~\ref{eq:latentnn_loss} dominates and the latent values remain close to the observations ($x_{\rm latent} \approx x_{\rm obs}$). At moderate SNR ($\mathrm{SNR}_x \approx 2$--3), the two loss terms compete more evenly, and weight decay tuning becomes important for LatentNN. The standard MLP, by contrast, fails regardless of weight decay---it lacks the mechanism to correct attenuation bias, and adjusting regularization only trades off between overfitting and underfitting without addressing the systematic bias.

LatentNN generalizes well to held-out test data. Weight decay regularization applied only to the network parameters $\boldsymbol{\theta}$ prevents the network from overfitting to the specific denoised distribution of the training latent values. The test set performance closely tracks the training set performance across all SNR values, indicating that the learned function $f_{\boldsymbol{\theta}}$ captures the true underlying relationship rather than artifacts of the training data.

\begin{figure*}[t]
\centering
\includegraphics[width=\textwidth]{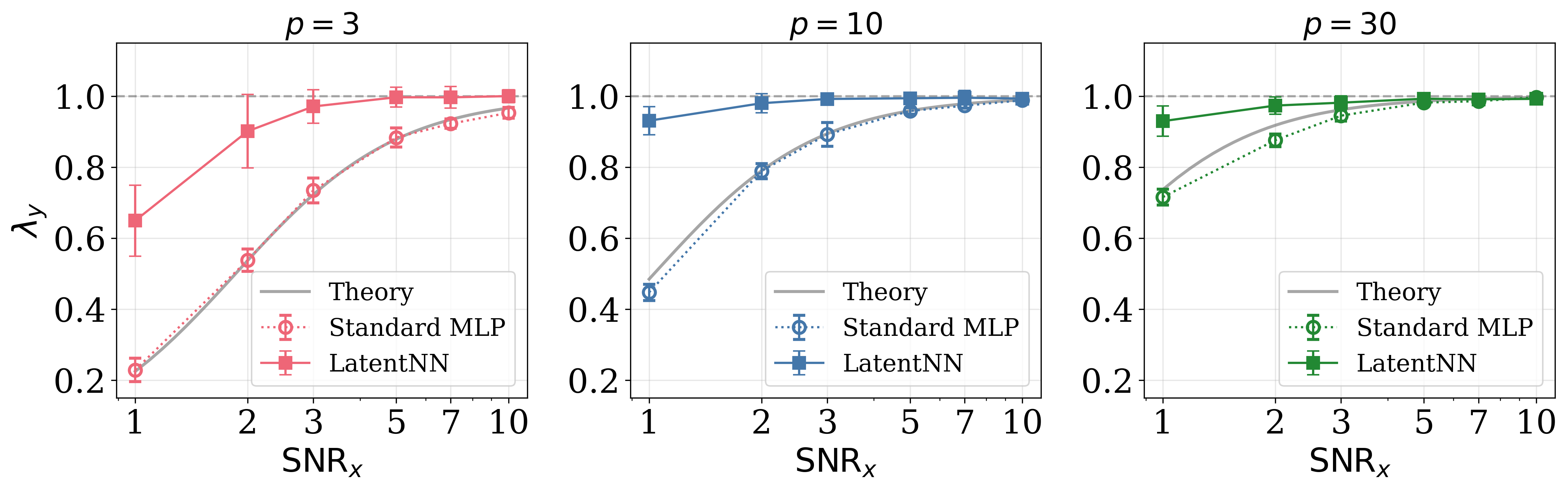}
\caption{Attenuation factor $\lambda_y$ versus $\mathrm{SNR}_x$ for correlated multivariate inputs with $p = 3, 10, 30$ dimensions (panels). Error bars show standard deviation over 8 runs. Black dashed line marks $\lambda_y = 1$ (no attenuation). Gray solid curve shows the analytic prediction for linear regression from Paper~I; standard MLP (open circles, dotted) closely follows this theoretical curve, confirming that model complexity does not mitigate the bias. With increasing $p$, attenuation is mitigated because independent noise in different dimensions must conspire coherently to displace observations along the signal direction. LatentNN (filled squares, solid) maintains $\lambda_y \approx 1$ for $p \geq 10$ at $\mathrm{SNR}_x \geq 2$, with mild degradation ($\lambda_y \approx 0.9$) at $\mathrm{SNR}_x = 1$. For $p=3$, LatentNN correction degrades at low SNR ($\lambda_y \approx 0.7$ at $\mathrm{SNR}_x = 1$), but still outperforms the standard MLP ($\lambda_y \approx 0.2$).}
\label{fig:multivariate_correlated}
\end{figure*}

\subsection{Multivariate Inputs}
\label{sec:multivariate}

Real astronomical applications involve multivariate inputs---for example, stellar spectra contain many flux measurements that serve as inputs to predict stellar parameters. The LatentNN formulation generalizes directly: the latent variables $\mathbf{X}_{\rm latent}$ form an $N \times p$ matrix, where $p$ is the number of input features. The loss function from Equation~\ref{eq:latentnn_loss} extends naturally to
\begin{align}
\mathcal{J} &= \sum_{i=1}^{N} \frac{(y_i - f_{\boldsymbol{\theta}}(\mathbf{X}_{{\rm latent},i}))^2}{\sigma_{y,i}^2} \nonumber \\
&\quad + \sum_{i=1}^{N} \sum_{j=1}^{p} \frac{(X_{{\rm obs},ij} - X_{{\rm latent},ij})^2}{\sigma_{x,ij}^2},
\label{eq:latentnn_multivariate}
\end{align}
where $\sigma_{y,i}$ is the output uncertainty for sample $i$ and $\sigma_{x,ij}$ is the input uncertainty for sample $i$ and dimension $j$. For simplicity, we assume homogeneous noise $\sigma_x$ and $\sigma_y$ across all samples and dimensions, which allows us to visualize results as a function of a single $\mathrm{SNR}_x$.

As shown in Paper I, for uncorrelated features each dimension attenuates independently, recovering the one-dimensional result. We therefore focus on features that are intrinsically correlated in the signal, which better represent astronomical data like stellar spectra where different pixels respond coherently to changes in stellar parameters. By ``correlated features'' we mean that the true input values $x_{{\rm true},j}$ are correlated with each other---not that the measurement errors are correlated. The measurement errors remain independent across features; it is the underlying signal that exhibits correlation structure.

To create such correlated features, we generate $N_{\rm train} = 1000$ training samples where all features derive from a single latent variable: $x_j = a_j z$ where $z \sim {\rm Uniform}(-0.5, 0.5)$ and $a_j$ varies linearly from 0.2 to 0.4 across dimensions. This construction ensures that the $p$ input features are perfectly correlated (all driven by the same underlying $z$), mimicking how spectral pixels respond coherently to changes in a stellar parameter. The true output is $y = \mathbf{x}^\top \mathbf{a}$, centered at zero. We fix $\mathrm{SNR}_y = 10$ as in the single-variate experiments. We use the same network architecture and training settings as Section~\ref{sec:nn_attenuation}, with the same three-fold data split (1000 train, 200 validation, 200 test) and weight decay grid search strategy. We test $p = 3, 10, 30$ dimensions at $\mathrm{SNR}_x$ values from 1 to 10, repeating each experiment 8 times.

Figure~\ref{fig:multivariate_correlated} shows the results. Paper I derived an analytic expression for the attenuation factor in correlated linear regression: $\lambda_y = \lambda_\beta = \sum_j a_j^2 / (1/\mathrm{SNR}_x^2 + \sum_j a_j^2)$. The gray curves show this theoretical prediction. Just as in the single-variate case, the standard MLP closely follows this linear regression theory curve, confirming that additional model complexity does not mitigate the fundamental attenuation bias. With correlated features, attenuation is weaker than the $p=1$ case because independent noise in different dimensions must conspire coherently to displace observations along the signal direction---a coincidence that becomes increasingly unlikely as $p$ grows. With increasing $p$, the standard MLP attenuation factor is mitigated compared to smaller $p$, but remains significant at lower SNR.

For LatentNN, the results reveal a non-monotonic relationship between dimensionality and correction difficulty. For $p \geq 10$, LatentNN maintains $\lambda_y \approx 1$ for $\mathrm{SNR}_x \geq 2$, with only mild degradation ($\lambda_y \approx 0.9$) at $\mathrm{SNR}_x = 1$. For $p=3$, however, correction degrades at low SNR: $\lambda_y \approx 0.7$ at $\mathrm{SNR}_x = 1$ and $\lambda_y \approx 0.9$ at $\mathrm{SNR}_x = 2$. Comparing to the single-variate case (Figure~\ref{fig:attenuation_comparison}), where $p=1$ achieves $\lambda_y \approx 1$ even at $\mathrm{SNR}_x = 1$, we see that intermediate dimensionality is harder than both low and high $p$.

This pattern reflects competing effects. Increasing $p$ generally helps because correlated features can lean on each other---the network learns their covariance structure and uses redundancy for more robust inference. However, this benefit only manifests when $p$ is large enough. For $p=1$, the problem is simple with a tractable one-dimensional posterior. For $p=3$, the posterior over latent values becomes more complex due to the increased dimensionality, yet there is insufficient redundancy among the three features to leverage. For $p \geq 10$, the higher-dimensional correlated structure provides enough constraints that features can collaboratively inform each other's latent values, enabling robust correction for $\mathrm{SNR}_x \geq 2$. Nevertheless, even in the challenging $p=3$ regime, LatentNN substantially outperforms the standard MLP: at $\mathrm{SNR}_x = 1$, LatentNN achieves $\lambda_y \approx 0.7$ compared to the standard MLP's $\lambda_y \approx 0.2$, more than tripling the recovered dynamic range.

\begin{figure*}[t]
\centering
\includegraphics[width=2\columnwidth]{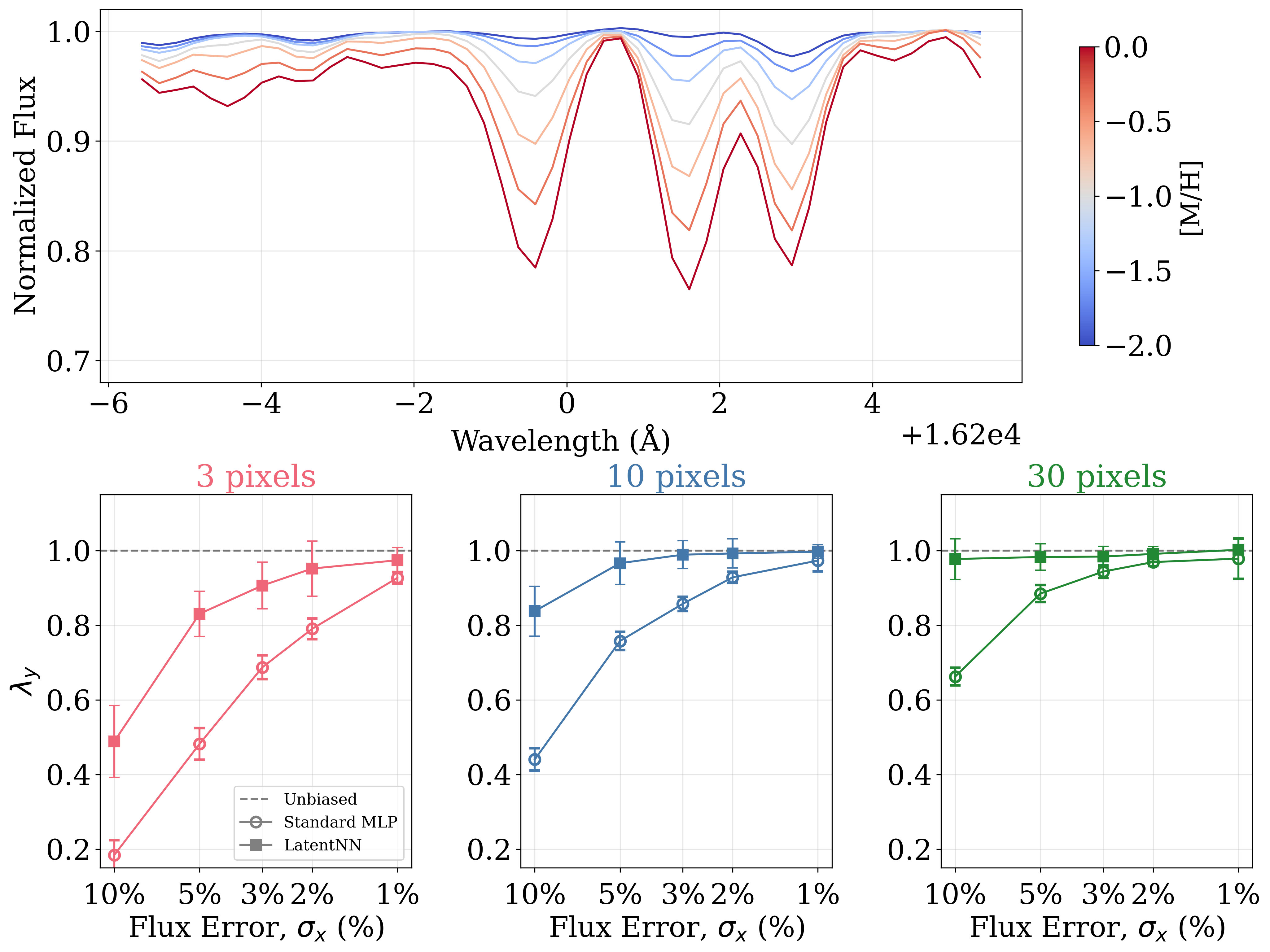}
\caption{Application to stellar spectra at $R \simeq 22500$ for inferring [M/H]. \textbf{Top:} Continuum-normalized flux variation with [M/H] from $-2.0$~dex (blue) to $0.0$~dex (red) in a metallicity-sensitive region around 16200~\AA. \textbf{Bottom:} Attenuation factor $\lambda_y$ versus flux error (\%) for 3, 10, and 30 contiguous pixels (panels). Error bars show standard deviation over 8 runs. The standard MLP (open circles) exhibits attenuation that worsens with fewer pixels and higher noise. LatentNN (filled squares) maintains $\lambda_y \gtrsim 0.95$ for 10 and 30 pixels across all noise levels. Even for 3 pixels where correction is more challenging, LatentNN also ameliorate the attenuation bias compared to the standard MLP (e.g., $\lambda_y \approx 0.5$ vs.\ $0.2$ at 10\% flux error).}
\label{fig:spectra}
\end{figure*}

\section{Application to Stellar Spectra}
\label{sec:spectra}

Having validated LatentNN on synthetic examples, we now apply it to a more realistic astronomical setting.

Stellar spectra represent the primary astronomical application motivating this work. To clarify the notation: in this spectroscopic context, the input vector $\mathbf{x}$ consists of flux measurements at different wavelength pixels, while the output $y$ is the stellar parameter we wish to infer (e.g., metallicity [M/H]). This follows our previous examples where $\mathbf{x}$ denotes the observed quantities with measurement error and $y$ is the label to be predicted. Data-driven models have been applied extensively to spectroscopic surveys \citep{BlancoCuaresma2014, Ness2015, Ting2017, Fabbro2018, Leung2019, Straumit2022, Xiang2022}, particularly for low-resolution and low-SNR observations where theoretical spectral models often fall short. These applications operate in the regime where attenuation bias is large. Even at high resolution ($R \sim 20000$), the intrinsic variation in continuum-normalized flux typically spans only $\sim$5--20\% ($\srange \approx 0.05$--0.2). 

For the weaker lines of $\srange \approx 0.05$, a spectrum with conventional SNR $= 20$ (defined relative to the normalized flux) has per-pixel uncertainty $\sigma_x = 0.05$, giving $\mathrm{SNR}_x = \srange/\sigma_x \approx 1$ in our definition; conventional SNR $= 100$ corresponds to $\mathrm{SNR}_x \approx 10$. We emphasize that what matters for attenuation bias is this ratio of signal range to measurement uncertainty, not the conventional spectroscopic SNR. Since the ratio $\srange/\sigma_x$ is often of order 10 or less for many spectral pixels in typical spectroscopic applications, they fall squarely in the regime where attenuation bias becomes important---especially for elements with weak or limited spectral features.

Real applications of LatentNN to survey data are underway, but this paper focuses on introducing the technique and establishing its theoretical foundation as an extension of Deming regression to neural networks. Here we focus on inferring overall metallicity [M/H] for clarity, but the formalism generalizes directly to multiple labels such as individual elemental abundances.

We use the pretrained synthetic spectral emulator with The Payne \citep{Ting2019} in The Payne github repository\footnote{https://github.com/tingyuansen/The\_Payne} to generate spectra at resolution $R \simeq 22500$, following Paper I. The underlying spectral model uses atomic and molecular line data from \citet{Kurucz1970, Kurucz1993, Kurucz2005, Kurucz2017}.\footnote{This emulator from the github repository of The Payne is trained on stellar models spanning [Fe/H] down to $-1.5$~dex. Our experiment extends to $-2.0$~dex, so we are extrapolating slightly. However, as seen in the top panel of Figure~\ref{fig:spectra}, the spectral variation remains smooth and physically reasonable. The exact accuracy of the spectral model is not critical to our conclusions about attenuation bias---the goal is to generate realistic correlated features mimicking the physical coupling between spectral pixels.} We focus on a metallicity-sensitive region around 16200~\AA.

Concretely, the experimental setup is as follows: we generate synthetic spectra for stars with varying [M/H] values, then train a neural network to predict [M/H] from the observed (noisy) flux values. The input $\mathbf{x}_i$ for sample $i$ is a vector of continuum-normalized flux measurements at $p$ wavelength pixels, and the output $y_i$ is the metallicity [M/H]. We add Gaussian noise to the flux values to simulate observational uncertainty, giving $\mathbf{x}_{{\rm obs},i} = \mathbf{x}_{{\rm true},i} + \boldsymbol{\delta}_x$ where $\boldsymbol{\delta}_x \sim \mathcal{N}(0, \sigma_x^2 \mathbf{I})$. The attenuation factor $\lambda_y$ is defined as before: the slope of the regression of predicted [M/H] against true [M/H] on a held-out test set.

To explore how the number of informative features affects attenuation and correction, we test three pixel configurations: 3, 10, and 30 contiguous pixels centered at 16200~\AA. These choices represent the range of spectral information available for different labels. While major elements like Fe have many strong lines, attenuation bias is most problematic for elements with only a few spectral features \citep[e.g., K, V; see fig.\ 7 and 8 of][]{Xiang2019}. In practice, even surveys with thousands of pixels often have only a small number of pixels that carry useful information for a given label.

We generate samples spanning [M/H] from $-2.0$ to $0.0$~dex, with fixed stellar parameters ($T_{\rm eff} = 4750$~K, $\log g = 2.5$, typical of red clump stars). All abundances scale with [M/H] for simplicity. We use the same three-fold data split as in the synthetic experiments (1000 train, 200 validation, 200 test) with hyperparameters selected on the validation set and final evaluation on the held-out test set. We add Gaussian noise with standard deviation $\sigma_x$ to each flux pixel and $\sigma_{\rm [M/H]} = 0.05$~dex to the metallicity labels, representing conservative uncertainties in training labels from surveys like APOGEE \citep{Holtzman2015, GarciaPerez2016}. We test flux error values $\sigma_x \in \{1\%, 2\%, 3\%, 5\%, 10\%\}$ of the normalized flux. We use the same network architecture and training settings as Section~\ref{sec:nn_attenuation}, with the same weight decay grid search strategy from Section~\ref{sec:results}. We repeat each experiment 8 times to estimate uncertainties.

Figure~\ref{fig:spectra} shows the results. The top panel illustrates how the spectral flux varies with [M/H] across the selected wavelength region, with color indicating metallicity from $-2.0$ (blue) to $0.0$~dex (red). Feature depths range from 5--20\%, but even features with 10--20\% total depth do not give $\mathrm{SNR}_x = 2$--4 at 5\% flux error as one might naively expect. This is because spectral features vary approximately quadratically with metallicity \citep{Rix2016}. In Paper I, we showed that the effective SNR is reduced by at least a factor of $n$ for polynomial features of degree $n$. For quadratic features, even $\srange = 0.1$ with $\sigma_x = 0.05$ yields an effective $\mathrm{SNR}_x \approx 1$ rather than 2. 

The bottom panels show $\lambda_y$ versus flux error for each pixel configuration. The behavior mirrors the synthetic multivariate experiments (Section~\ref{sec:multivariate}): the standard MLP exhibits attenuation that worsens with fewer pixels and higher noise, while LatentNN provides robust correction when sufficient pixels are available. For 10 and 30 pixels, LatentNN maintains $\lambda_y \gtrsim 0.95$ across all tested noise levels. For 3 pixels, however, correction degrades at high noise: $\lambda_y \approx 0.5$ at 10\% flux error and $\lambda_y \approx 0.8$ at 5\% flux error. This matches the intermediate-dimensionality challenge discussed in Section~\ref{sec:multivariate}---with few pixels, the posterior over latent values is complex yet there is insufficient redundancy to leverage. Even so, LatentNN outperforms the standard MLP across all configurations: at 10\% flux error with 3 pixels, LatentNN achieves $\lambda_y \approx 0.5$ compared to the standard MLP's $\lambda_y \approx 0.2$.

For spectral features with $\srange \approx 0.10$--0.20 (strong lines), flux errors of $\sigma_x \lesssim 3$--5\% yield effective $\mathrm{SNR}_x \gtrsim 2$ after accounting for the factor of $\sim$2 reduction from quadratic features---consistent with our findings throughout that LatentNN works well for $\mathrm{SNR}_x \geq 2$. For weaker features with $\srange \approx 0.05$, correspondingly lower flux errors ($\sigma_x \lesssim 1$--2\%) are needed. At higher noise levels with few informative pixels, careful hyperparameter tuning becomes essential, and users should expect some residual bias.

\section{Discussion}
\label{sec:discussion}

\subsection{Implications for Spectroscopic Surveys}

The approach of transferring stellar parameters and abundances from high-resolution surveys like APOGEE \citep{Holtzman2015, Majewski2017} to lower-resolution surveys like LAMOST \citep{Luo2015, Ho2017, Xiang2019}, DESI \citep{Zhang2024}, and Gaia XP \citep{Andrae2023, Andrae2023b, Zhang2023_XP, Fallows2024, Khalatyan2024, Li2024, Hattori2025, Yang2025} has become standard practice---but these applications operate precisely in the regime where attenuation bias is severe. 

Even at APOGEE resolution, attenuation bias is non-negligible at moderate SNR. For lower-resolution surveys where individual spectral features are blended and per-pixel $\srange$ is smaller, the ratio $\sx/\srange$ is larger and attenuation bias is more severe. Crucially, $\srange$ scales approximately proportionally with spectral resolution $R$: a survey like LAMOST ($R \sim 1800$) has $\srange$ roughly 10 times smaller than APOGEE ($R \sim 22500$). This means that even at high conventional SNR $\sim 100$ ($\sigma_x \sim 1\%$), the effective $\mathrm{SNR}_x$ remains in the regime where attenuation bias is significant.

Our spectroscopic experiments (Figure~\ref{fig:spectra}) quantify the severity. For 10 informative pixels at 10\% flux error (conventional SNR~$\approx 10$), the standard MLP yields $\lambda_y \approx 0.5$. Assuming training data centered around [M/H]~$= -1$, a halo star with true [M/H]~$= -2$ would be predicted as [M/H]~$\approx -1.5$, potentially misclassified as a thick-disk star. Even at 5\% flux error (conventional SNR~$\approx 20$), we find $\lambda_y \approx 0.7$ for 10 pixels---a star at [M/H]~$= -2$ would be predicted as [M/H]~$\approx -1.7$. While this might seem acceptable for elements with many features, it is problematic for elements with fewer than 10 informative spectral lines. For 3 pixels at 10\% error, $\lambda_y$ drops to $\approx 0.2$, meaning a [M/H]~$= -2$ star would be predicted as [M/H]~$\approx -1.2$.

This has practical consequences---metal-poor stars are systematically predicted to be more metal-rich, making the rarest objects harder to identify. This bias directly affects the inferred metallicity distribution function (MDF), particularly its metal-poor tail, which is crucial for constraining the formation history of the proto-Milky Way \citep{Rix2022, Chen2024, Chen2025}. If attenuation bias compresses the metallicity scale, the true slope of the metal-poor MDF tail would be steeper than observed, with implications for inferred gas masses, star formation efficiencies, and inflow histories in galactic chemical evolution models. Stellar ages suffer from compression at both ends, with the oldest stars underestimated and the youngest overestimated. Distances are similarly affected, leading to systematic errors in the inferred geometry and dynamics of the Milky Way. Obtaining better training labels does not solve the problem because the bias arises from noisy inputs, not labels.

When information is concentrated in sparse spectral features such as individual absorption lines, attenuation is most severe. Consider an element like K or V with only 2--3 usable lines, compared to Fe with dozens of strong features: K and V abundances are effectively in the low-dimensional regime where $\sigma_x/\srange$ is large, while Fe benefits from the high-dimensional correlated regime. This pattern---elements with fewer lines showing larger systematic errors at extreme values---has been observed in practice \citep{Xiang2019, Zhang2024} but is often attributed to model limitations or data quality issues. Attenuation bias is a fundamental contributor, and the LatentNN framework provides a path toward unbiased parameter estimation for data-driven deep learning application to spectroscopic data. 

\subsection{Connection to Hierarchical Bayesian Modeling}
\label{sec:connection}

The errors-in-variables problem has a long history in statistics \citep{Deming1943, Fuller1987}. Deming regression and orthogonal regression \citep{GolubVanLoan1980, VanHuffelVandewalle1991} treat the true input values as latent variables to be estimated alongside the regression coefficients---but this approach has seen limited application in astronomy \citep[but see][]{Kelly2007},  in part because astronomical relationships are often nonlinear and proper feature engineering can be difficult. Standard neural network frameworks do not include errors-in-variables handling by default because they arose in domains like computer vision where input noise is negligible. LatentNN extends the classical latent variable approach to neural networks, providing a principled framework for astronomical applications where input uncertainties are non-negligible.

Our LatentNN formulation can be understood as the maximum a posteriori (MAP) limit of a hierarchical Bayesian model. In the full Bayesian formulation, the true values $\xtrue$ are latent variables with prior distributions centered on the observed values $\xobs$, and we would infer the posterior distribution over both the model parameters $\boldsymbol{\theta}$ and all the latent values. The LatentNN optimization finds the mode of this posterior. Neural networks provide a flexible function approximator that avoids the need for explicit feature engineering, while the latent variable framework provides the statistical foundation for handling input uncertainties.

This formulation provides the foundation for understanding why correlated features can lean on each other, as discussed in Section~\ref{sec:multivariate}. When the model learns the correlated structure of the inputs, it can use this structure to constrain the latent values: each feature's latent value is informed not only by its own observation but also by what the learned model expects given the other features.

It is important to note that the latent values $\xlat$ should not be interpreted strictly as ``denoised'' inputs. In high-dimensional settings with large $p$, the posterior distribution over $\xlat$ can be broad and complex---many combinations of latent values may be consistent with the observed data and the learned model. What matters for our purposes is that the network parameters $\boldsymbol{\theta}$ remain well-constrained: even when individual latent values have substantial uncertainty, the function $f_{\boldsymbol{\theta}}$ that maps inputs to outputs is well-defined. The optimization finds a configuration where the network generalizes correctly to new data, achieving unbiased predictions ($\lambda_y \approx 1$) even if the individual latent values do not perfectly recover the true inputs. This is the sense in which LatentNN corrects attenuation bias---through the learned function rather than through perfect recovery of latent values. At low SNR or intermediate dimensionality where the posterior is particularly complex (the $p=3$ regime in Figure~\ref{fig:multivariate_correlated}), case-by-case hyperparameter tuning may be needed, but a systematic exploration of such regimes is beyond the scope of this paper.

It is worth comparing LatentNN with alternative strategies that might address input uncertainties. Training with noise-augmented data \citep{Fabbro2018}---resampling input noise at each epoch---improves robustness but does not correct the systematic bias. Each augmented sample is still drawn from $\mathcal{N}(x_{\rm obs}, \sigma_x^2)$, so the expected input remains $x_{\rm obs}$; the optimization still minimizes loss conditioned on noisy inputs, and the attenuation factor $\lambda_y$ is unchanged. Predicting uncertainties as additional outputs \citep{Leung2019} quantifies scatter but does not remove bias in the predictive mean. Denoising autoencoders learn a mapping from noisy to clean inputs \citep{Vojtekova2021, Gheller2022, Pal2024}, but the denoised values themselves can exhibit attenuation bias when the autoencoder is trained on noisy data. Bayesian neural networks place distributions over the network \emph{weights} to capture epistemic uncertainty, but do not model uncertainty in the \emph{inputs}. The posterior predictive mean $\mathbb{E}_{\boldsymbol{\theta}}[f_{\boldsymbol{\theta}}(\xobs)]$ is still conditioned on the noisy $\xobs$ rather than the unknown $\xtrue$, so attenuation bias in the mean prediction is unchanged by construction---marginalizing over weight uncertainty does not correct a bias that arises from the inputs. Variational autoencoders \citep{KingmaWelling2013} learn latent representations, but their latent space is a lower-dimensional learned embedding rather than a corrected version of the physical input space with known noise properties. Gaussian process regression can in principle incorporate input noise, but standard GP scaling ($\mathcal{O}(N^3)$ in time) makes this impractical for the dataset sizes typical in astronomy. The approach most closely related to LatentNN is full hierarchical Bayesian inference with latent true inputs \citep{Kelly2007}, which provides full posterior distributions and is in principle the gold standard. LatentNN can be understood as the MAP limit of this framework (Section~\ref{sec:connection}), trading posterior characterization for scalability by leveraging standard neural network optimization infrastructure. LatentNN addresses the root cause by explicitly modeling the relationship between observed and true input values, providing a scalable optimization-based alternative.

\subsection{Generalizations}

The core idea of maximizing a likelihood with latent variables extends to many settings beyond the homoscedastic Gaussian case demonstrated here. As shown in Equation~\ref{eq:latentnn_multivariate}, heteroscedastic uncertainties are handled by per-sample weights. If measurement errors are correlated across pixels, as is common in spectra due to sky subtraction and continuum normalization, the latent likelihood term incorporates the noise covariance matrix $\mathbf{\Sigma}_{x,i}$:
\begin{align}
\mathcal{J} &= \sum_{i=1}^{N} \frac{1}{\sigma_{y,i}^2} (y_i - f_{\boldsymbol{\theta}}(\mathbf{x}_{{\rm latent},i}))^2 \nonumber \\
&\quad + \sum_{i=1}^{N} (\mathbf{x}_{{\rm obs},i} - \mathbf{x}_{{\rm latent},i})^\top \mathbf{\Sigma}_{x,i}^{-1} (\mathbf{x}_{{\rm obs},i} - \mathbf{x}_{{\rm latent},i}).
\end{align}
For Poisson noise, outliers, or other non-Gaussian errors, the Gaussian likelihood can be replaced with the appropriate distribution, giving $\mathcal{J} = -\log p(\xobs | \xlat) - \log p(\yobs | f_{\boldsymbol{\theta}}(\xlat))$ for whatever form $p$ takes.

For classification tasks with noisy inputs, the same latent variable approach applies. The prediction term becomes cross-entropy loss instead of mean squared error, while the latent likelihood term remains unchanged:
\begin{align}
\mathcal{J} &= {\rm CrossEntropy}(f_{\boldsymbol{\theta}}(\xlat), y) \nonumber \\
&\quad + \sum_{i=1}^{N} (\mathbf{x}_{{\rm obs},i} - \mathbf{x}_{{\rm latent},i})^\top \mathbf{\Sigma}_{x,i}^{-1} (\mathbf{x}_{{\rm obs},i} - \mathbf{x}_{{\rm latent},i}).
\end{align}
Extension to other neural network architectures requires only that the network be differentiable with respect to its inputs. Convolutional neural networks for imaging data, transformers for sequential data, or any other architecture can be used by simply treating the input as learnable parameters with an appropriate latent likelihood term.

\subsection{Limitations}

The method requires specifying both input uncertainties $\sx$ (or $\sigma_{x,i}$ in the heteroscedastic case) and output uncertainties $\sy$. In practice, $\sx$ may need to be estimated based on the noise model of the instrument, while $\sy$ reflects label uncertainties or, when labels are precise, a nominal value that sets the relative weighting in the loss function. Misspecified $\sx$ or $\sy$ leads to suboptimal correction: underestimating $\sx$ causes undercorrection, while overestimating it can cause overcorrection. A possible remedy is adding jitter terms to be optimized along with the other parameters.

LatentNN performs optimization rather than sampling. We obtain point estimates of $x_{{\rm latent},i}$ and $\boldsymbol{\theta}$, but characterizing the full posterior over the high-dimensional space $(\boldsymbol{\theta}, \xlat)$ is challenging. Full Bayesian treatment of neural networks is already difficult, and adding the latent variables makes it more so. Approximate Bayesian approaches such as dropout \citep{Hinton2012, Srivastava2014} or variational inference \citep{KingmaWelling2013} may provide practical uncertainty estimates. 

The increased number of parameters also leads to a more complex loss landscape. The optimization may converge to different solutions from different initializations, particularly at low SNR where the problem is ill-conditioned. Careful hyperparameter tuning, especially of the weight decay parameter as discussed in Section~\ref{sec:results}, is often necessary.

In practice, however, the results are not highly sensitive to the weight decay value. In the one-dimensional experiments at $\mathrm{SNR}_x = 2$, sweeping weight decay over three orders of magnitude ($10^{-4}$ to $10^{-1}$) yields $\lambda_y$ values between 0.94 and 0.98---a variation of only $\sim$4\%. For the multivariate case with $p = 10$ correlated features, the variation is even smaller ($\sim$2\%). The dominant source of variability is not hyperparameter choice but rather run-to-run scatter from random initialization, which increases at low SNR: the standard deviation across 8 random seeds grows from $\sim$0.01 at $\mathrm{SNR}_x = 10$ to $\sim$0.09 at $\mathrm{SNR}_x = 1$. We recommend searching over a logarithmically spaced grid (e.g., $10^{-4}, 3 \times 10^{-4}, 10^{-3}, \ldots, 10^{-1}$) and selecting the value that yields $\lambda_y$ closest to unity on a held-out validation set. When true labels are not available for computing $\lambda_y$ directly on new data, a practical alternative is to monitor the balance between the prediction loss and latent likelihood terms during training: a well-tuned weight decay produces a model where neither term dominates excessively at convergence.

The computational cost scales as $\mathcal{O}(N \times p)$, introducing $N$ additional parameters per input dimension. For full spectra with thousands of pixels and training sets of tens of thousands of stars, this becomes a consideration. To give concrete estimates: for an APOGEE-like training set with $N = 50{,}000$ and $p = 100$ informative pixels, LatentNN introduces $5 \times 10^6$ additional latent parameters ($\sim$20\,MB in float32), well within the capacity of a single modern GPU. Using the full spectral vector ($p \sim 7000$) increases this to $\sim$3.5$\times 10^8$ parameters ($\sim$1.4\,GB), which remains feasible on GPUs with $\geq$16\,GB memory. For larger training sets ($N \sim 10^6$) at full spectral dimensionality, the latent parameters alone would require $\sim$28\,GB, pushing the limits of single-GPU memory but accessible with high-memory GPUs or multi-GPU configurations. In practice, restricting to informative pixels ($p \sim 30$--$100$) keeps the problem comfortably tractable even for large $N$. Scalability requires care, particularly for high-dimensional inference problems. Possible strategies include using subsets of informative pixels. Dimensionality reduction before LatentNN is less straightforward because the noise properties in the reduced space may not be well characterized. In summary, LatentNN is most directly applicable to problems with moderate training set sizes ($N \sim 10^3$--$10^5$) and a manageable number of informative input features ($p \sim 10$--$100$), in the signal-to-noise regime where attenuation bias is significant ($\mathrm{SNR}_x \lesssim 10$) but correction remains reliable ($\mathrm{SNR}_x \gtrsim 2$). In astronomy, this encompasses many spectroscopic applications, particularly for elements with limited spectral features, as well as photometric and time-series problems where input uncertainties are non-negligible. Extension to larger datasets or higher-dimensional inputs is feasible with appropriate feature selection, though full-dimensional application to very large training sets remains a computational challenge.

Despite these limitations, the method provides a practical solution for a fundamental problem for which existing remedies---such as full Bayesian inference over latent variables---have not yet been demonstrated in the neural network context and face substantial computational challenges when scaling to large datasets.

\section{Conclusions}
\label{sec:conclusions}

Neural networks have become essential tools for astronomical inference, from deriving stellar parameters from spectra to estimating photometric redshifts from imaging data. These applications often involve low-SNR observations where measurement uncertainties are non-negligible. When inputs are noisy, attenuation bias becomes a fundamental concern.

This effect, well known in the statistics literature for linear regression \citep{Fuller1987}, extends to neural networks. Neural networks trained on noisy inputs suffer from the same bias as linear regression, with coefficient slopes biased toward zero by a factor that scales with $\sx/\srange$. Increasing model complexity does not mitigate the effect. This issue has received little attention in the machine learning literature because most applications involve high-SNR inputs where the effect is negligible. Astronomy is different, and as neural networks become increasingly central to astronomical inference, classical statistical considerations such as errors-in-variables become important.

Attenuation bias in linear regression is well understood, and solutions such as Deming regression have existed for decades. These methods treat the true input values as latent variables to be estimated alongside the model parameters. We show that this approach generalizes from linear to nonlinear models. By simultaneously inferring what the true inputs were, the model learns the correct input-output relationship rather than one diluted by measurement noise.

We propose LatentNN as a practical implementation of this idea. The method adds a learnable latent value for each training input, constrained by a latent likelihood term to stay close to the observed value, and jointly optimizes these latent values with the network parameters. This is a simple modification that can be applied to any differentiable architecture. We validated LatentNN on one-dimensional regression, where it matches Deming regression for the linear case, on multivariate inputs with varying correlation structure, and on realistic spectral applications using mock data with varying numbers of informative pixels. LatentNN effectively mitigates attenuation bias, particularly for $\mathrm{SNR}_x \gtrsim 2$, with the degree of correction depending on input dimensionality and correlation structure.

For astronomy, this work provides a framework for improved inference with neural networks in the low-SNR regime that characterizes much of astronomical data. The method is relevant for spectroscopic surveys, photometric applications, and any data-driven model where inputs carry non-negligible measurement uncertainty. The biases we address affect measurements at the extremes of parameter space, which are often the most scientifically interesting.

Neural network modeling is a powerful tool, but it is not a panacea. Neural networks inherit the same limitations as classical statistical methods when those limitations arise from the data rather than the model. Attenuation bias is one such limitation. Recognizing when classical statistical insights apply to modern machine learning methods, and adapting solutions accordingly, is essential for reliable scientific inference in astronomical research.

\section*{Code Availability}

We provide a reference implementation of LatentNN at \url{https://github.com/tingyuansen/LatentNN}. The code demonstrates the method for simple fully-connected architectures as presented in this paper. The approach is general and can be adapted to any differentiable architecture by treating the input as learnable parameters with an appropriate latent likelihood term.

\begin{acknowledgments}

I thank David Weinberg and Hans-Walter Rix for useful discussion and comments on the manuscript. The original idea about errors-in-variables stemmed from my teaching of Astron 5550 (Advanced Data Analysis), during which I decided to discuss attenuation bias as explored in Paper~I. While preparing for that lecture, I was inspired to read more into the errors-in-variables literature, also suggested by an anonymous referee during Paper~I. Further discussion with Jeffrey Newman and Brett Andrews during a colloquium at UPitt inspired the extension of this work to deep learning, which ultimately led to this paper.

This research was supported by NSF Grant AST-2406729 and a Humboldt Research Award from the Alexander von Humboldt Foundation.

Claude 4.5 was used for assistance with code development and manuscript copy-editing. The scientific content, analysis, and conclusions are the author’s, and all LLM-assisted changes were verified by the author.

\end{acknowledgments}

\appendix

\section{Derivation of Attenuation Bias}
\label{app:attenuation}

Consider a true linear relationship $y_{\rm true} = \beta x_{\rm true}$. Our observations are
\begin{align}
x_{\rm obs} &= x_{\rm true} + \delta_x, \\
y_{\rm obs} &= \beta x_{\rm true} + \delta_y,
\end{align}
where $\delta_x$ and $\delta_y$ are independent measurement errors with variances $\sx^2$ and $\sy^2$.

The standard least squares estimator is $\hat{\beta} = {\rm Cov}(x_{\rm obs}, y_{\rm obs})/{\rm Var}(x_{\rm obs})$. Expanding these terms:
\begin{align}
{\rm Cov}(x_{\rm obs}, y_{\rm obs}) &= {\rm Cov}(x_{\rm true} + \delta_x, \beta x_{\rm true} + \delta_y) \\
&= \beta\, {\rm Var}(x_{\rm true}) = \beta \srange^2,
\end{align}
since measurement errors are independent of true values and of each other. For the denominator:
\begin{equation}
{\rm Var}(x_{\rm obs}) = {\rm Var}(x_{\rm true}) + {\rm Var}(\delta_x) = \srange^2 + \sx^2.
\end{equation}

The expected value of the standard estimator is therefore
\begin{equation}
\mathbb{E}[\hat{\beta}] = \beta \frac{\srange^2}{\srange^2 + \sx^2} = \beta \cdot \lambda_\beta,
\end{equation}
where $\lambda_\beta = 1/(1 + (\sx/\srange)^2) < 1$ is the attenuation factor. The estimated slope is systematically biased toward zero.

\section{Derivation of Deming Regression}
\label{app:deming}

To correct attenuation bias, we treat the true values $\{x_{{\rm true},i}\}$ as latent variables to be estimated alongside $\beta$. The joint likelihood, conditioned on the true values, is
\begin{equation}
\mathcal{L} = \prod_{i=1}^N \mathcal{N}(x_{{\rm obs},i} | x_{{\rm true},i}, \sx^2) \cdot \mathcal{N}(y_{{\rm obs},i} | \beta x_{{\rm true},i}, \sy^2).
\end{equation}

Expanding the normal distributions and taking the negative log-likelihood (dropping constants):
\begin{equation}
E(\{x_{{\rm true},i}\}, \beta) = \sum_{i=1}^{N} \left[ \frac{(x_{{\rm obs},i} - x_{{\rm true},i})^2}{2\sx^2} + \frac{(y_{{\rm obs},i} - \beta x_{{\rm true},i})^2}{2\sy^2} \right].
\end{equation}

\begin{figure}[t]
    \centering
    \includegraphics[width=0.8\columnwidth]{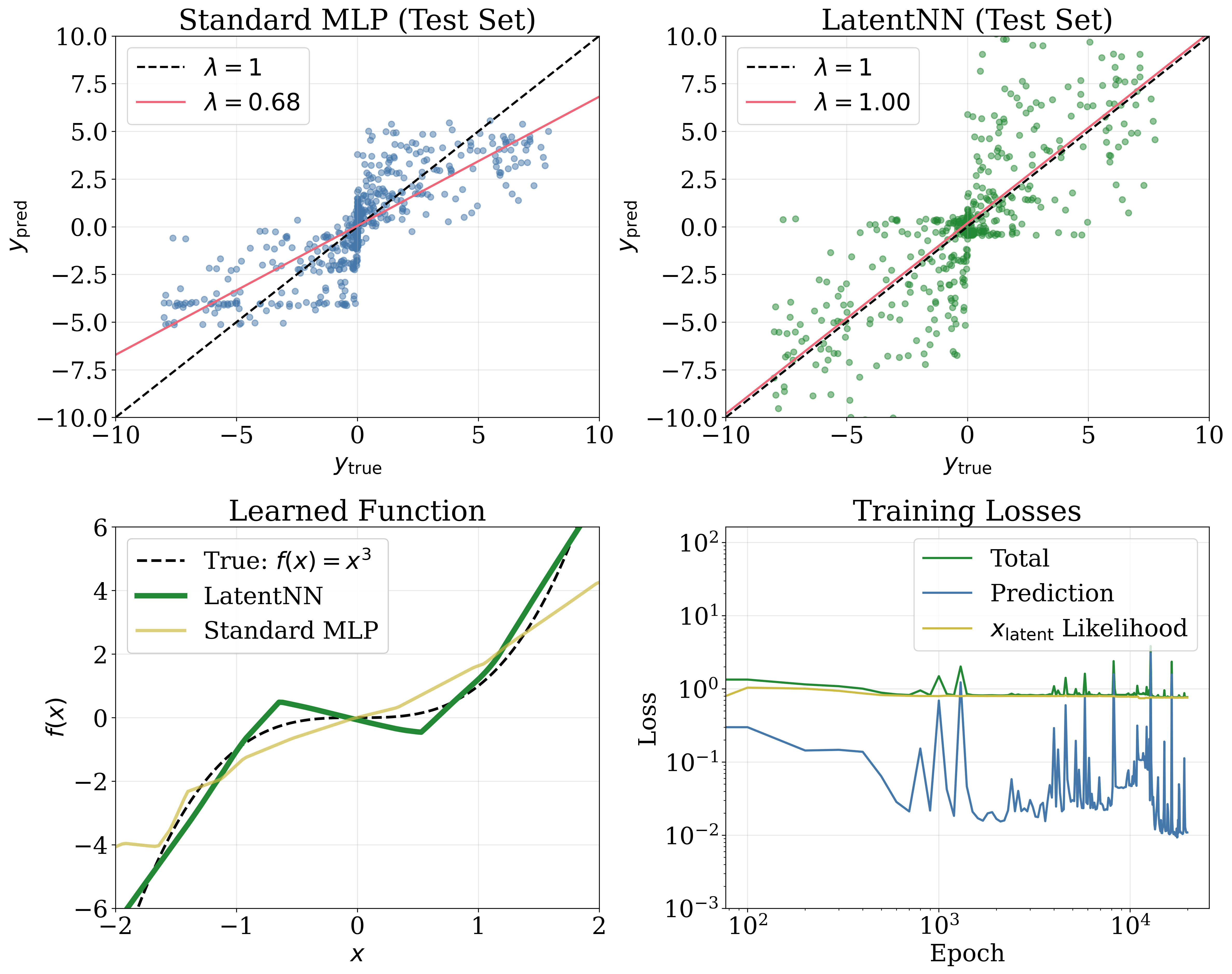}
    \caption{LatentNN applied to a nonlinear function $f(x) = x^3$ at $\mathrm{SNR}_x = 2$. \textbf{Top:} Predicted versus true $y$ on the test set for standard MLP (left, $\lambda_y \approx 0.7$) and LatentNN (right, $\lambda_y \approx 1$). \textbf{Bottom left:} Learned functions compared to the true cubic. The standard MLP learns a flattened curve, while LatentNN recovers the correct shape. \textbf{Bottom right:} LatentNN training losses showing the trade-off between prediction accuracy and latent variable likelihood.}
    \label{fig:cubic}
\end{figure}

For a fixed $\beta$, each $x_{{\rm true},i}$ appears only in the $i$th term, so we can optimize them independently. Taking the derivative and setting it to zero:
\begin{equation}
\frac{\partial E}{\partial x_{{\rm true},i}} = -\frac{x_{{\rm obs},i} - x_{{\rm true},i}}{\sx^2} - \frac{\beta(y_{{\rm obs},i} - \beta x_{{\rm true},i})}{\sy^2} = 0.
\end{equation}

Rearranging and letting $r = \sy^2/\sx^2$ denote the error variance ratio:
\begin{equation}
x_{{\rm true},i} = \frac{r\, x_{{\rm obs},i} + \beta y_{{\rm obs},i}}{r + \beta^2}.
\end{equation}

This can be rewritten as a weighted average:
\begin{equation}
x_{{\rm true},i} = w_x \cdot x_{{\rm obs},i} + w_y \cdot \frac{y_{{\rm obs},i}}{\beta},
\end{equation}
where $w_x = r/(r + \beta^2)$ and $w_y = \beta^2/(r + \beta^2)$ sum to unity. When $r \gg 1$ (precise $x$ measurements), $w_x \approx 1$ and we trust $x_{\rm obs}$. When $r \ll 1$ (precise $y$ measurements), $w_y \approx 1$ and we infer $x$ from $y/\beta$.

Taking the derivative of $E$ with respect to $\beta$, substituting the expression for $x_{{\rm true},i}$, and using the summary statistics $S_{xx} = \sum_i (x_{{\rm obs},i} - \bar{x})^2$, $S_{yy} = \sum_i (y_{{\rm obs},i} - \bar{y})^2$, and $S_{xy} = \sum_i (x_{{\rm obs},i} - \bar{x})(y_{{\rm obs},i} - \bar{y})$, we obtain a quadratic equation:
\begin{equation}
S_{xy} \beta^2 - (S_{yy} - r S_{xx})\beta - r S_{xy} = 0.
\end{equation}

Applying the quadratic formula and taking the positive root:
\begin{equation}
\hat{\beta} = \frac{S_{yy} - r S_{xx} + \sqrt{(S_{yy} - r S_{xx})^2 + 4 r S_{xy}^2}}{2 S_{xy}}.
\end{equation}

In the limit $\sx \rightarrow 0$ (equivalently $r \rightarrow \infty$), this reduces to $\hat{\beta} \rightarrow S_{xy}/S_{xx}$, the ordinary least squares estimator.

\section{Extension to Nonlinear Functions}
\label{app:nonlinear}

The main text focuses on linear functions because they permit direct comparison with the theoretical attenuation factor $\lambda_\beta = 1/(1 + (\sx/\srange)^2)$. However, the errors-in-variables problem and the LatentNN correction are more general and apply to arbitrary nonlinear functions.

To demonstrate this, we test LatentNN on a cubic polynomial $f(x) = x^3$. We use a larger network capacity (2 layers $\times$ 256 hidden units) to ensure the network can represent the nonlinear function accurately. Figure~\ref{fig:cubic} shows results at $\mathrm{SNR}_x = 2$, a challenging regime where noise equals 50\% of the signal range. The same weight decay hyperparameter search procedure from Section~\ref{sec:results} is applied.

The standard MLP exhibits severe attenuation, with $\lambda_y \approx 0.65$---even worse than the linear case at the same SNR, as derived in Paper I. This occurs because measurement errors in $x$ are amplified by the nonlinear function: an error $\delta_x$ produces a prediction error of approximately $f'(x) \delta_x$, which varies across the input domain. Near the edges of the training range where $|f'(x)| = 3x^2$ is largest, the effective noise is amplified.

LatentNN corrects this bias. The bottom-left panel shows the learned functions: the standard MLP learns a flattened version of the cubic, while LatentNN recovers the true function shape. This demonstrates that the latent variable framework extends naturally beyond linear regression to nonlinear function approximation.

\bibliographystyle{aasjournal}
\bibliography{manuscript}

@ARTICLE{Ting2017,
       author = {{Ting}, Yuan-Sen and {Rix}, Hans-Walter and {Conroy}, Charlie and {Ho}, Anna Y.~Q. and {Lin}, Jane},
        title = "{Measuring 14 Elemental Abundances with R = 1800 LAMOST Spectra}",
      journal = {\apjl},
         year = 2017,
        month = nov,
       volume = {849},
       number = {1},
          eid = {L9},
        pages = {L9},
          doi = {10.3847/2041-8213/aa921c},
archivePrefix = {arXiv},
       eprint = {1708.01758},
 primaryClass = {astro-ph.SR},
       adsurl = {https://ui.adsabs.harvard.edu/abs/2017ApJ...849L...9T}
}

@ARTICLE{Rix2016,
       author = {{Rix}, Hans-Walter and {Ting}, Yuan-Sen and {Conroy}, Charlie and {Hogg}, David W.},
        title = "{Constructing Polynomial Spectral Models for Stars}",
      journal = {\apjl},
         year = 2016,
        month = aug,
       volume = {826},
       number = {2},
          eid = {L25},
        pages = {L25},
          doi = {10.3847/2041-8205/826/2/L25},
archivePrefix = {arXiv},
       eprint = {1603.06574},
 primaryClass = {astro-ph.SR},
       adsurl = {https://ui.adsabs.harvard.edu/abs/2016ApJ...826L..25R}
}

@ARTICLE{Andrae2023,
       author = {{Andrae}, R. and {Fouesneau}, M. and {Sordo}, R. and {Bailer-Jones}, C.~A.~L. and {Dharmawardena}, T.~E. and {Rybizki}, J. and {De Angeli}, F. and {Lindstr{\o}m}, H.~E.~P. and {Marshall}, D.~J. and {Drimmel}, R. and {Korn}, A.~J. and {Soubiran}, C. and {Brouillet}, N. and {Casamiquela}, L. and {Rix}, H.-W. and {Abreu Aramburu}, A. and {{\'A}lvarez}, M.~A. and {Bakker}, J. and {Bellas-Velidis}, I. and {Bijaoui}, A. and {Brugaletta}, E. and {Burlacu}, A. and {Carballo}, R. and {Chaoul}, L. and {Chiavassa}, A. and {Contursi}, G. and {Cooper}, W.~J. and {Creevey}, O.~L. and {Dafonte}, C. and {Dapergolas}, A. and {de Laverny}, P. and {Delchambre}, L. and {Demouchy}, C. and {Edvardsson}, B. and {Fr{\'e}mat}, Y. and {Garabato}, D. and {Garc{\'\i}a-Lario}, P. and {Garc{\'\i}a-Torres}, M. and {Gavel}, A. and {Gomez}, A. and {Gonz{\'a}lez-Santamar{\'\i}a}, I. and {Hatzidimitriou}, D. and {Heiter}, U. and {Jean-Antoine Piccolo}, A. and {Kontizas}, M. and {Kordopatis}, G. and {Lanzafame}, A.~C. and {Lebreton}, Y. and {Licata}, E.~L. and {Livanou}, E. and {Lobel}, A. and {Lorca}, A. and {Magdaleno Romeo}, A. and {Manteiga}, M. and {Marocco}, F. and {Mary}, N. and {Nicolas}, C. and {Ordenovic}, C. and {Pailler}, F. and {Palicio}, P.~A. and {Pallas-Quintela}, L. and {Panem}, C. and {Pichon}, B. and {Poggio}, E. and {Recio-Blanco}, A. and {Riclet}, F. and {Robin}, C. and {Santove{\~n}a}, R. and {Sarro}, L.~M. and {Schultheis}, M.~S. and {Segol}, M. and {Silvelo}, A. and {Slezak}, I. and {Smart}, R.~L. and {S{\"u}veges}, M. and {Th{\'e}venin}, F. and {Torralba Elipe}, G. and {Ulla}, A. and {Utrilla}, E. and {Vallenari}, A. and {van Dillen}, E. and {Zhao}, H. and {Zorec}, J.},
        title = "{Gaia Data Release 3. Analysis of the Gaia BP/RP spectra using the General Stellar Parameterizer from Photometry}",
      journal = {\aap},
         year = 2023,
        month = jun,
       volume = {674},
          eid = {A27},
        pages = {A27},
          doi = {10.1051/0004-6361/202243462},
archivePrefix = {arXiv},
       eprint = {2206.06138},
 primaryClass = {astro-ph.SR},
       adsurl = {https://ui.adsabs.harvard.edu/abs/2023A&A...674A..27A}
}

@BOOK{Bishop2006,
  author = {Bishop, Christopher M.},
  title = {Pattern Recognition and Machine Learning (Information Science and Statistics)},
  year = {2006},
  isbn = {0387310738},
  publisher = {Springer-Verlag},
  address = {Berlin, Heidelberg}
}

@ARTICLE{BlancoCuaresma2014,
       author = {{Blanco-Cuaresma}, S. and {Soubiran}, C. and {Heiter}, U. and {Jofr{\'e}}, P.},
        title = "{Determining stellar atmospheric parameters and chemical abundances of FGK stars with iSpec}",
      journal = {\aap},
         year = 2014,
        month = sep,
       volume = {569},
          eid = {A111},
        pages = {A111},
          doi = {10.1051/0004-6361/201423945},
archivePrefix = {arXiv},
       eprint = {1407.2608},
 primaryClass = {astro-ph.IM},
       adsurl = {https://ui.adsabs.harvard.edu/abs/2014A&A...569A.111B}
}

@ARTICLE{Brown2020,
       author = {{Brown}, Tom B. and {Mann}, Benjamin and {Ryder}, Nick and {Subbiah}, Melanie and {Kaplan}, Jared and {Dhariwal}, Prafulla and {Neelakantan}, Arvind and {Shyam}, Pranav and {Sastry}, Girish and {Askell}, Amanda and {Agarwal}, Sandhini and {Herbert-Voss}, Ariel and {Krueger}, Gretchen and {Henighan}, Tom and {Child}, Rewon and {Ramesh}, Aditya and {Ziegler}, Daniel M. and {Wu}, Jeffrey and {Winter}, Clemens and {Hesse}, Christopher and {Chen}, Mark and {Sigler}, Eric and {Litwin}, Mateusz and {Gray}, Scott and {Chess}, Benjamin and {Clark}, Jack and {Berner}, Christopher and {McCandlish}, Sam and {Radford}, Alec and {Sutskever}, Ilya and {Amodei}, Dario},
        title = "{Language Models are Few-Shot Learners}",
      journal = {arXiv e-prints},
         year = 2020,
        month = may,
          eid = {arXiv:2005.14165},
        pages = {arXiv:2005.14165},
          doi = {10.48550/arXiv.2005.14165},
archivePrefix = {arXiv},
       eprint = {2005.14165},
 primaryClass = {cs.CL},
       adsurl = {https://ui.adsabs.harvard.edu/abs/2020arXiv200514165B}
}

@book{Carroll1995,
       author = {{Carroll}, R. J. and {Ruppert}, D. and {Stefanski}, L. A.},
        title = "{Measurement Error in Non-linear Models}",
    publisher = {Chapman and Hall},
     address = {New York},
         year = 1995
}

@book{Carroll2006,
  author = {Carroll, R. J. and Ruppert, D. and Stefanski, L. A. and Crainiceanu, C. M.},
  title = {Measurement error in nonlinear models: A modern perspective, Second edition},
  publisher = {Chapman and Hall},
  year = {2006}
}

@ARTICLE{Cranmer2020a,
       author = {{Cranmer}, Kyle and {Brehmer}, Johann and {Louppe}, Gilles},
        title = "{The frontier of simulation-based inference}",
      journal = {Proceedings of the National Academy of Science},
         year = 2020,
        month = dec,
       volume = {117},
       number = {48},
        pages = {30055-30062},
          doi = {10.1073/pnas.1912789117},
archivePrefix = {arXiv},
       eprint = {1911.01429},
 primaryClass = {stat.ML},
       adsurl = {https://ui.adsabs.harvard.edu/abs/2020PNAS..11730055C}
}

@ARTICLE{Cybenko1989,
  author = {{Cybenko}, G.},
  title = "{Approximation by superpositions of a sigmoidal function}",
  journal = {Mathematics of Control, Signals, and Systems},
  year = 1989,
  month = dec,
  volume = {2},
  number = {4},
  pages = {303-314},
  doi = {10.1007/BF02551274},
  adsurl = {https://ui.adsabs.harvard.edu/abs/1989MCSS....2..303C}
}

@book{Deming1943,
  author = {Deming, W. E.},
  title = {Statistical adjustment of data},
  publisher = {Wiley},
  year = {1943}
}

@ARTICLE{Devlin2019,
       author = {{Devlin}, Jacob and {Chang}, Ming-Wei and {Lee}, Kenton and {Toutanova}, Kristina},
        title = "{BERT: Pre-training of Deep Bidirectional Transformers for Language Understanding}",
      journal = {arXiv e-prints},
         year = 2018,
        month = oct,
          eid = {arXiv:1810.04805},
        pages = {arXiv:1810.04805},
          doi = {10.48550/arXiv.1810.04805},
archivePrefix = {arXiv},
       eprint = {1810.04805},
 primaryClass = {cs.CL},
       adsurl = {https://ui.adsabs.harvard.edu/abs/2018arXiv181004805D}
}

@ARTICLE{Fabbro2018,
       author = {{Fabbro}, S. and {Venn}, K.~A. and {O'Briain}, T. and {Bialek}, S. and {Kielty}, C.~L. and {Jahandar}, F. and {Monty}, S.},
        title = "{An application of deep learning in the analysis of stellar spectra}",
      journal = {\mnras},
         year = 2018,
        month = apr,
       volume = {475},
       number = {3},
        pages = {2978-2993},
          doi = {10.1093/mnras/stx3298},
archivePrefix = {arXiv},
       eprint = {1709.09182},
 primaryClass = {astro-ph.IM},
       adsurl = {https://ui.adsabs.harvard.edu/abs/2018MNRAS.475.2978F}
}

@article{Frost2002,
    author = {Frost, Chris and Thompson, Simon G.},
    title = {Correcting for Regression Dilution Bias: Comparison of Methods for a Single Predictor Variable},
    journal = {Journal of the Royal Statistical Society Series A: Statistics in Society},
    volume = {163},
    number = {2},
    pages = {173-189},
    year = {2002},
    month = {01},
    issn = {0964-1998},
    doi = {10.1111/1467-985X.00164},
    url = {https://doi.org/10.1111/1467-985X.00164},
    eprint = {https://academic.oup.com/jrsssa/article-pdf/163/2/173/49761025/jrsssa\_163\_2\_173.pdf},
}

@book{Fuller1987,
       author = {{Fuller}, W. A.},
        title = "{Measurement Error Models}",
    publisher = {Wiley},
     address = {New York},
         year = 1987
}

@ARTICLE{GarciaPerez2016,
       author = {{Garc{\'\i}a P{\'e}rez}, Ana E. and {Allende Prieto}, Carlos and {Holtzman}, Jon A. and {Shetrone}, Matthew and {M{\'e}sz{\'a}ros}, Szabolcs and {Bizyaev}, Dmitry and {Carrera}, Ricardo and {Cunha}, Katia and {Garc{\'\i}a-Hern{\'a}ndez}, D.~A. and {Johnson}, Jennifer A. and {Majewski}, Steven R. and {Nidever}, David L. and {Schiavon}, Ricardo P. and {Shane}, Neville and {Smith}, Verne V. and {Sobeck}, Jennifer and {Troup}, Nicholas and {Zamora}, Olga and {Weinberg}, David H. and {Bovy}, Jo and {Eisenstein}, Daniel J. and {Feuillet}, Diane and {Frinchaboy}, Peter M. and {Hayden}, Michael R. and {Hearty}, Fred R. and {Nguyen}, Duy C. and {O'Connell}, Robert W. and {Pinsonneault}, Marc H. and {Wilson}, John C. and {Zasowski}, Gail},
        title = "{ASPCAP: The APOGEE Stellar Parameter and Chemical Abundances Pipeline}",
      journal = {\aj},
         year = 2016,
        month = jun,
       volume = {151},
       number = {6},
          eid = {144},
        pages = {144},
          doi = {10.3847/0004-6256/151/6/144},
archivePrefix = {arXiv},
       eprint = {1510.07635},
 primaryClass = {astro-ph.SR},
       adsurl = {https://ui.adsabs.harvard.edu/abs/2016AJ....151..144G}
}

@book{Gelman2013,
  author = {Gelman, A. and Carlin, J. B. and Stern, H. S. and Dunson, D. B. and Vehtari, A. and Rubin, D. B.},
  title = {Bayesian data analysis (3rd ed.)},
  publisher = {Chapman and Hall},
  year = {2013}
}

@article{GolubVanLoan1980,
  author = {Golub, G. H. and Van Loan, C. F.},
  title = {An analysis of the total least squares problem},
  journal = {SIAM Journal on Numerical Analysis},
  volume = {17},
  number = {6},
  pages = {883--893},
  year = {1980}
}

@ARTICLE{Hinton2012,
       author = {{Hinton}, Geoffrey E. and {Srivastava}, Nitish and {Krizhevsky}, Alex and {Sutskever}, Ilya and {Salakhutdinov}, Ruslan R.},
        title = "{Improving neural networks by preventing co-adaptation of feature detectors}",
      journal = {arXiv e-prints},
         year = 2012,
        month = jul,
          eid = {arXiv:1207.0580},
        pages = {arXiv:1207.0580},
          doi = {10.48550/arXiv.1207.0580},
archivePrefix = {arXiv},
       eprint = {1207.0580},
 primaryClass = {cs.NE},
       adsurl = {https://ui.adsabs.harvard.edu/abs/2012arXiv1207.0580H}
}

@ARTICLE{Ho2017,
       author = {{Ho}, Anna Y.~Q. and {Ness}, Melissa K. and {Hogg}, David W. and {Rix}, Hans-Walter and {Liu}, Chao and {Yang}, Fan and {Zhang}, Yong and {Hou}, Yonghui and {Wang}, Yuefei},
        title = "{Label Transfer from APOGEE to LAMOST: Precise Stellar Parameters for 450,000 LAMOST Giants}",
      journal = {\apj},
         year = 2017,
        month = feb,
       volume = {836},
       number = {1},
          eid = {5},
        pages = {5},
          doi = {10.3847/1538-4357/836/1/5},
archivePrefix = {arXiv},
       eprint = {1602.00303},
 primaryClass = {astro-ph.SR},
       adsurl = {https://ui.adsabs.harvard.edu/abs/2017ApJ...836....5H}
}

@ARTICLE{Holtzman2015,
       author = {{Holtzman}, Jon A. and {Shetrone}, Matthew and {Johnson}, Jennifer A. and {Allende Prieto}, Carlos and {Anders}, Friedrich and {Andrews}, Brett and {Beers}, Timothy C. and {Bizyaev}, Dmitry and {Blanton}, Michael R. and {Bovy}, Jo and {Carrera}, Ricardo and {Chojnowski}, S. Drew and {Cunha}, Katia and {Eisenstein}, Daniel J. and {Feuillet}, Diane and {Frinchaboy}, Peter M. and {Galbraith-Frew}, Jessica and {Garc{\'\i}a P{\'e}rez}, Ana E. and {Garc{\'\i}a-Hern{\'a}ndez}, D.~A. and {Hasselquist}, Sten and {Hayden}, Michael R. and {Hearty}, Fred R. and {Ivans}, Inese and {Majewski}, Steven R. and {Martell}, Sarah and {Meszaros}, Szabolcs and {Muna}, Demitri and {Nidever}, David and {Nguyen}, Duy Cuong and {O'Connell}, Robert W. and {Pan}, Kaike and {Pinsonneault}, Marc and {Robin}, Annie C. and {Schiavon}, Ricardo P. and {Shane}, Neville and {Sobeck}, Jennifer and {Smith}, Verne V. and {Troup}, Nicholas and {Weinberg}, David H. and {Wilson}, John C. and {Wood-Vasey}, W.~M. and {Zamora}, Olga and {Zasowski}, Gail},
        title = "{Abundances, Stellar Parameters, and Spectra from the SDSS-III/APOGEE Survey}",
      journal = {\aj},
         year = 2015,
        month = nov,
       volume = {150},
       number = {5},
          eid = {148},
        pages = {148},
          doi = {10.1088/0004-6256/150/5/148},
archivePrefix = {arXiv},
       eprint = {1501.04110},
 primaryClass = {astro-ph.GA},
       adsurl = {https://ui.adsabs.harvard.edu/abs/2015AJ....150..148H}
}

@ARTICLE{Hon2021,
       author = {{Hon}, Marc and {Huber}, Daniel and {Kuszlewicz}, James S. and {Stello}, Dennis and {Sharma}, Sanjib and {Tayar}, Jamie and {Zinn}, Joel C. and {Vrard}, Mathieu and {Pinsonneault}, Marc H.},
        title = "{A ``Quick Look'' at All-sky Galactic Archeology with TESS: 158,000 Oscillating Red Giants from the MIT Quick-look Pipeline}",
      journal = {\apj},
         year = 2021,
        month = oct,
       volume = {919},
       number = {2},
          eid = {131},
        pages = {131},
          doi = {10.3847/1538-4357/ac14b1},
archivePrefix = {arXiv},
       eprint = {2108.01241},
 primaryClass = {astro-ph.SR},
       adsurl = {https://ui.adsabs.harvard.edu/abs/2021ApJ...919..131H}
}

@article{HornikStinchcombeWhite1989,
  author  = {Hornik, K. and Stinchcombe, M. and White, H.},
  title   = {Multilayer feedforward networks are universal approximators},
  journal = {Neural Networks},
  volume  = {2},
  number  = {5},
  pages   = {359-366},
  year    = {1989}
}

@ARTICLE{Hoyle2016,
       author = {{Hoyle}, B.},
        title = "{Measuring photometric redshifts using galaxy images and Deep Neural Networks}",
      journal = {Astronomy and Computing},
         year = 2016,
        month = jul,
       volume = {16},
        pages = {34-40},
          doi = {10.1016/j.ascom.2016.03.006},
archivePrefix = {arXiv},
       eprint = {1504.07255},
 primaryClass = {astro-ph.IM},
       adsurl = {https://ui.adsabs.harvard.edu/abs/2016A&C....16...34H}
}

@ARTICLE{Huertas-Company2023,
       author = {{Huertas-Company}, M. and {Lanusse}, F.},
        title = "{The Dawes Review 10: The impact of deep learning for the analysis of galaxy surveys}",
      journal = {\pasa},
         year = 2023,
        month = jan,
       volume = {40},
          eid = {e001},
        pages = {e001},
          doi = {10.1017/pasa.2022.55},
archivePrefix = {arXiv},
       eprint = {2210.01813},
 primaryClass = {astro-ph.IM},
       adsurl = {https://ui.adsabs.harvard.edu/abs/2023PASA...40....1H}
}

@ARTICLE{Kelly2007,
       author = {{Kelly}, Brandon C.},
        title = "{Some Aspects of Measurement Error in Linear Regression of Astronomical Data}",
      journal = {\apj},
         year = 2007,
        month = aug,
       volume = {665},
       number = {2},
        pages = {1489-1506},
          doi = {10.1086/519947},
archivePrefix = {arXiv},
       eprint = {0705.2774},
 primaryClass = {astro-ph},
       adsurl = {https://ui.adsabs.harvard.edu/abs/2007ApJ...665.1489K}
}

@ARTICLE{Khalatyan2024,
       author = {{Khalatyan}, A. and {Anders}, F. and {Chiappini}, C. and {Queiroz}, A.~B.~A. and {Nepal}, S. and {dal Ponte}, M. and {Jordi}, C. and {Guiglion}, G. and {Valentini}, M. and {Torralba Elipe}, G. and {Steinmetz}, M. and {Pantaleoni-Gonz{\'a}lez}, M. and {Malhotra}, S. and {Jim{\'e}nez-Arranz}, {\'O}. and {Enke}, H. and {Casamiquela}, L. and {Ard{\`e}vol}, J.},
        title = "{Transferring spectroscopic stellar labels to 217 million Gaia DR3 XP stars with SHBoost}",
      journal = {\aap},
         year = 2024,
        month = nov,
       volume = {691},
          eid = {A98},
        pages = {A98},
          doi = {10.1051/0004-6361/202451427},
archivePrefix = {arXiv},
       eprint = {2407.06963},
 primaryClass = {astro-ph.SR},
       adsurl = {https://ui.adsabs.harvard.edu/abs/2024A&A...691A..98K}
}

@ARTICLE{KingmaWelling2013,
       author = {{Kingma}, Diederik P and {Welling}, Max},
        title = "{Auto-Encoding Variational Bayes}",
      journal = {arXiv e-prints},
         year = 2013,
        month = dec,
          eid = {arXiv:1312.6114},
        pages = {arXiv:1312.6114},
          doi = {10.48550/arXiv.1312.6114},
archivePrefix = {arXiv},
       eprint = {1312.6114},
 primaryClass = {stat.ML},
       adsurl = {https://ui.adsabs.harvard.edu/abs/2013arXiv1312.6114K}
}

@ARTICLE{Kurucz1970,
       author = {{Kurucz}, R.~L.},
        title = "{Atlas: a Computer Program for Calculating Model Stellar Atmospheres}",
      journal = {SAO Special Report},
         year = 1970,
        month = feb,
       volume = {309},
       adsurl = {https://ui.adsabs.harvard.edu/abs/1970SAOSR.309.....K}
}

@BOOK{Kurucz1993,
       author = {{Kurucz}, Robert L.},
        title = "{SYNTHE spectrum synthesis programs and line data}",
         year = 1993,
       adsurl = {https://ui.adsabs.harvard.edu/abs/1993sssp.book.....K}
}

@ARTICLE{Kurucz2005,
       author = {{Kurucz}, R.~L.},
        title = "{ATLAS12, SYNTHE, ATLAS9, WIDTH9, et cetera}",
      journal = {Memorie della Societa Astronomica Italiana Supplementi},
         year = 2005,
       volume = {8},
        pages = {14},
       adsurl = {https://ui.adsabs.harvard.edu/abs/2005MSAIS...8...14K}
}

@MISC{Kurucz2017,
       author = {{Kurucz}, R.~L.},
        title = "{ATLAS9: Model atmosphere program with opacity distribution functions}",
 howpublished = {Astrophysics Source Code Library},
         year = 2017,
        month = oct,
archivePrefix = {ascl},
       eprint = {1710.017},
       adsurl = {https://ui.adsabs.harvard.edu/abs/2017ascl.soft10017K}
}

@article{Lecun1998,
  author  = {LeCun, Y. and Bottou, L. and Bengio, Y. and Haffner, P.},
  title   = {Gradient-based learning applied to document recognition},
  journal = {Proceedings of the IEEE},
  volume  = {86},
  number  = {11},
  pages   = {2278-2324},
  year    = {1998}
}

@ARTICLE{Leung2019,
       author = {{Leung}, Henry W. and {Bovy}, Jo},
        title = "{Deep learning of multi-element abundances from high-resolution spectroscopic data}",
      journal = {\mnras},
         year = 2019,
        month = mar,
       volume = {483},
       number = {3},
        pages = {3255-3277},
          doi = {10.1093/mnras/sty3217},
archivePrefix = {arXiv},
       eprint = {1808.04428},
 primaryClass = {astro-ph.GA},
       adsurl = {https://ui.adsabs.harvard.edu/abs/2019MNRAS.483.3255L}
}

@ARTICLE{Li2022_GaZNet,
       author = {{Li}, Rui and {Napolitano}, Nicola R. and {Feng}, Haicheng and {Li}, Ran and {Amaro}, Valeria and {Xie}, Linghua and {Tortora}, Crescenzo and {Bilicki}, Maciej and {Brescia}, Massimo and {Cavuoti}, Stefano and {Radovich}, Mario},
        title = "{Galaxy morphoto-Z with neural Networks (GaZNets). I. Optimized accuracy and outlier fraction from imaging and photometry}",
      journal = {\aap},
         year = 2022,
        month = oct,
       volume = {666},
          eid = {A85},
        pages = {A85},
          doi = {10.1051/0004-6361/202244081},
archivePrefix = {arXiv},
       eprint = {2205.10720},
 primaryClass = {astro-ph.GA},
       adsurl = {https://ui.adsabs.harvard.edu/abs/2022A&A...666A..85L}
}

@ARTICLE{Li2024,
       author = {{Li}, Jiadong and {Wong}, Kaze W.~K. and {Hogg}, David W. and {Rix}, Hans-Walter and {Chandra}, Vedant},
        title = "{AspGap: Augmented Stellar Parameters and Abundances for 37 Million Red Giant Branch Stars from Gaia XP Low-resolution Spectra}",
      journal = {\apjs},
         year = 2024,
        month = may,
       volume = {272},
       number = {1},
          eid = {2},
        pages = {2},
          doi = {10.3847/1538-4365/ad2b4d},
archivePrefix = {arXiv},
       eprint = {2309.14294},
 primaryClass = {astro-ph.SR},
       adsurl = {https://ui.adsabs.harvard.edu/abs/2024ApJS..272....2L}
}

@ARTICLE{Lin2022,
       author = {{Lin}, Q. and {Fouchez}, D. and {Pasquet}, J. and {Treyer}, M. and {Ait Ouahmed}, R. and {Arnouts}, S. and {Ilbert}, O.},
        title = "{Photometric redshift estimation with convolutional neural networks and galaxy images: Case study of resolving biases in data-driven methods}",
      journal = {\aap},
         year = 2022,
        month = jun,
       volume = {662},
          eid = {A36},
        pages = {A36},
          doi = {10.1051/0004-6361/202142751},
archivePrefix = {arXiv},
       eprint = {2202.09964},
 primaryClass = {astro-ph.IM},
       adsurl = {https://ui.adsabs.harvard.edu/abs/2022A&A...662A..36L}
}

@INPROCEEDINGS{Loredo2004,
       author = {{Loredo}, Thomas J.},
        title = "{Accounting for Source Uncertainties in Analyses of Astronomical Survey Data}",
    booktitle = {Bayesian Inference and Maximum Entropy Methods in Science and Engineering: 24th International Workshop on Bayesian Inference and Maximum Entropy Methods in Science and Engineering},
         year = 2004,
       editor = {{Fischer}, Rainer and {Preuss}, Roland and {Toussaint}, Udo Von},
       series = {American Institute of Physics Conference Series},
       volume = {735},
        month = nov,
    publisher = {AIP},
        pages = {195-206},
          doi = {10.1063/1.1835214},
archivePrefix = {arXiv},
       eprint = {astro-ph/0409387},
 primaryClass = {astro-ph},
       adsurl = {https://ui.adsabs.harvard.edu/abs/2004AIPC..735..195L}
}

@ARTICLE{Majewski2017,
       author = {{Majewski}, Steven R. and {Schiavon}, Ricardo P. and {Frinchaboy}, Peter M. and {Allende Prieto}, Carlos and {Barkhouser}, Robert and {Bizyaev}, Dmitry and {Blank}, Basil and {Brunner}, Sophia and {Burton}, Adam and {Carrera}, Ricardo and {Chojnowski}, S. Drew and {Cunha}, K{\'a}tia and {Epstein}, Courtney and {Fitzgerald}, Greg and {Garc{\'\i}a P{\'e}rez}, Ana E. and {Hearty}, Fred R. and {Henderson}, Chuck and {Holtzman}, Jon A. and {Johnson}, Jennifer A. and {Lam}, Charles R. and {Lawler}, James E. and {Maseman}, Paul and {M{\'e}sz{\'a}ros}, Szabolcs and {Nelson}, Matthew and {Nguyen}, Duy Coung and {Nidever}, David L. and {Pinsonneault}, Marc and {Shetrone}, Matthew and {Smee}, Stephen and {Smith}, Verne V. and {Stolberg}, Todd and {Skrutskie}, Michael F. and {Walker}, Eric and {Wilson}, John C. and {Zasowski}, Gail and {Anders}, Friedrich and {Basu}, Sarbani and {Beland}, Stephane and {Blanton}, Michael R. and {Bovy}, Jo and {Brownstein}, Joel R. and {Carlberg}, Joleen and {Chaplin}, William and {Chiappini}, Cristina and {Eisenstein}, Daniel J. and {Elsworth}, Yvonne and {Feuillet}, Diane and {Fleming}, Scott W. and {Galbraith-Frew}, Jessica and {Garc{\'\i}a}, Rafael A. and {Garc{\'\i}a-Hern{\'a}ndez}, D. An{\'\i}bal and {Gillespie}, Bruce A. and {Girardi}, L{\'e}o and {Gunn}, James E. and {Hasselquist}, Sten and {Hayden}, Michael R. and {Hekker}, Saskia and {Ivans}, Inese and {Kinemuchi}, Karen and {Klaene}, Mark and {Mahadevan}, Suvrath and {Mathur}, Savita and {Mosser}, Beno{\^\i}t and {Muna}, Demitri and {Munn}, Jeffrey A. and {Nichol}, Robert C. and {O'Connell}, Robert W. and {Parejko}, John K. and {Robin}, A.~C. and {Rocha-Pinto}, Helio and {Schultheis}, Matthias and {Serenelli}, Aldo M. and {Shane}, Neville and {Silva Aguirre}, Victor and {Sobeck}, Jennifer S. and {Thompson}, Benjamin and {Troup}, Nicholas W. and {Weinberg}, David H. and {Zamora}, Olga},
        title = "{The Apache Point Observatory Galactic Evolution Experiment (APOGEE)}",
      journal = {\aj},
         year = 2017,
        month = sep,
       volume = {154},
       number = {3},
          eid = {94},
        pages = {94},
          doi = {10.3847/1538-3881/aa784d},
archivePrefix = {arXiv},
       eprint = {1509.05420},
 primaryClass = {astro-ph.IM},
       adsurl = {https://ui.adsabs.harvard.edu/abs/2017AJ....154...94M}
}

@ARTICLE{Ness2015,
       author = {{Ness}, M. and {Hogg}, David W. and {Rix}, H.-W. and {Ho}, Anna. Y.~Q. and {Zasowski}, G.},
        title = "{The Cannon: A data-driven approach to Stellar Label Determination}",
      journal = {\apj},
         year = 2015,
        month = jul,
       volume = {808},
       number = {1},
          eid = {16},
        pages = {16},
          doi = {10.1088/0004-637X/808/1/16},
archivePrefix = {arXiv},
       eprint = {1501.07604},
 primaryClass = {astro-ph.SR},
       adsurl = {https://ui.adsabs.harvard.edu/abs/2015ApJ...808...16N}
}

@ARTICLE{Pal2024,
       author = {{P{\'a}l}, Bal{\'a}zs and {Dobos}, L{\'a}szl{\'o}},
        title = "{Denoising Medium Resolution Stellar Spectra With Neural Networks}",
      journal = {Astronomische Nachrichten},
         year = 2024,
        month = nov,
       volume = {345},
          eid = {e20240049},
        pages = {e20240049},
          doi = {10.1002/asna.20240049},
archivePrefix = {arXiv},
       eprint = {2409.11625},
 primaryClass = {astro-ph.IM},
       adsurl = {https://ui.adsabs.harvard.edu/abs/2024AN....34540049P}
}

@article{RumelhartHintonWilliams1986,
  author  = {Rumelhart, D. E. and Hinton, G. E. and Williams, R. J.},
  title   = {Learning representations by back-propagating errors},
  journal = {Nature},
  volume  = {323},
  pages   = {533--536},
  year    = {1986}
}

@ARTICLE{Simonyan2014,
       author = {{Simonyan}, Karen and {Zisserman}, Andrew},
        title = "{Very Deep Convolutional Networks for Large-Scale Image Recognition}",
      journal = {arXiv e-prints},
         year = 2014,
        month = sep,
          eid = {arXiv:1409.1556},
        pages = {arXiv:1409.1556},
          doi = {10.48550/arXiv.1409.1556},
archivePrefix = {arXiv},
       eprint = {1409.1556},
 primaryClass = {cs.CV},
       adsurl = {https://ui.adsabs.harvard.edu/abs/2014arXiv1409.1556S}
}

@ARTICLE{Spearman1904,
       author = {{Spearman}, C.},
        title = "{The proof and measurement of association between two things}",
      journal = {The American Journal of Psychology},
         year = 1904,
       volume = {15},
       number = {1},
        pages = {72-101},
          doi = {10.2307/1412159}
}

@article{Srivastava2014,
  author  = {Srivastava, N. and Hinton, G. E. and Krizhevsky, A. and Sutskever, I. and Salakhutdinov, R. R.},
  title   = {Dropout: a simple way to prevent neural networks from overfitting},
  journal = {Journal of Machine Learning Research},
  volume  = {15},
  number  = {56},
  pages   = {1929--1958},
  year    = {2014}
}

@ARTICLE{Straumit2022,
       author = {{Straumit}, Ilya and {Tkachenko}, Andrew and {Gebruers}, Sarah and {Audenaert}, Jeroen and {Xiang}, Maosheng and {Zari}, Eleonora and {Aerts}, Conny and {Johnson}, Jennifer A. and {Kollmeier}, Juna A. and {Rix}, Hans-Walter and {Beaton}, Rachael L. and {Van Saders}, Jennifer L. and {Teske}, Johanna and {Roman-Lopes}, Alexandre and {Ting}, Yuan-Sen and {Rom{\'a}n-Z{\'u}{\~n}iga}, Carlos G.},
        title = "{Zeta-Payne: A Fully Automated Spectrum Analysis Algorithm for the Milky Way Mapper Program of the SDSS-V Survey}",
      journal = {\aj},
         year = 2022,
        month = may,
       volume = {163},
       number = {5},
          eid = {236},
        pages = {236},
          doi = {10.3847/1538-3881/ac5f49},
archivePrefix = {arXiv},
       eprint = {2203.14538},
 primaryClass = {astro-ph.IM},
       adsurl = {https://ui.adsabs.harvard.edu/abs/2022AJ....163..236S}
}

@ARTICLE{Szegedy2014,
       author = {{Szegedy}, Christian and {Liu}, Wei and {Jia}, Yangqing and {Sermanet}, Pierre and {Reed}, Scott and {Anguelov}, Dragomir and {Erhan}, Dumitru and {Vanhoucke}, Vincent and {Rabinovich}, Andrew},
        title = "{Going Deeper with Convolutions}",
      journal = {arXiv e-prints},
         year = 2014,
        month = sep,
          eid = {arXiv:1409.4842},
        pages = {arXiv:1409.4842},
          doi = {10.48550/arXiv.1409.4842},
archivePrefix = {arXiv},
       eprint = {1409.4842},
 primaryClass = {cs.CV},
       adsurl = {https://ui.adsabs.harvard.edu/abs/2014arXiv1409.4842S}
}

@ARTICLE{Ting2019,
       author = {{Ting}, Yuan-Sen and {Conroy}, Charlie and {Rix}, Hans-Walter and {Cargile}, Phillip},
        title = "{The Payne: Self-consistent ab initio Fitting of Stellar Spectra}",
      journal = {\apj},
         year = 2019,
        month = jul,
       volume = {879},
       number = {2},
          eid = {69},
        pages = {69},
          doi = {10.3847/1538-4357/ab2331},
archivePrefix = {arXiv},
       eprint = {1804.01530},
 primaryClass = {astro-ph.SR},
       adsurl = {https://ui.adsabs.harvard.edu/abs/2019ApJ...879...69T}
}

@ARTICLE{Ting2025,
       author = {{Ting}, Yuan-Sen},
        title = "{Deep Learning in Astrophysics}",
      journal = {arXiv e-prints},
         year = 2025,
        month = oct,
          eid = {arXiv:2510.10713},
        pages = {arXiv:2510.10713},
          doi = {10.48550/arXiv.2510.10713},
archivePrefix = {arXiv},
       eprint = {2510.10713},
 primaryClass = {astro-ph.IM},
       adsurl = {https://ui.adsabs.harvard.edu/abs/2025arXiv251010713T}
}

@ARTICLE{Ting2025_PaperI,
       author = {{Ting}, Yuan-Sen},
        title = "{Why Machine Learning Models Systematically Underestimate Extreme Values}",
      journal = {The Open Journal of Astrophysics},
         year = 2025,
        month = jul,
       volume = {8},
          eid = {95},
        pages = {95},
          doi = {10.33232/001c.142224},
archivePrefix = {arXiv},
       eprint = {2412.05806},
 primaryClass = {astro-ph.IM},
       adsurl = {https://ui.adsabs.harvard.edu/abs/2025OJAp....8E..95T}
}

@ARTICLE{Ting2025b,
       author = {{Ting}, Yuan-Sen},
        title = "{Statistical Machine Learning for Astronomy -- A Textbook}",
      journal = {arXiv e-prints},
         year = 2025,
        month = jun,
          eid = {arXiv:2506.12230},
        pages = {arXiv:2506.12230},
          doi = {10.48550/arXiv.2506.12230},
archivePrefix = {arXiv},
       eprint = {2506.12230},
 primaryClass = {astro-ph.IM},
       adsurl = {https://ui.adsabs.harvard.edu/abs/2025arXiv250612230T}
}

@book{VanHuffelVandewalle1991,
  author = {Van Huffel, S. and Vandewalle, J.},
  title = {The total least squares problem: Computational aspects and analysis},
  publisher = {Society for Industrial and Applied Mathematics},
  year = {1991}
}

@ARTICLE{Vojtekova2021,
       author = {{Vojtekova}, Antonia and {Lieu}, Maggie and {Valtchanov}, Ivan and {Altieri}, Bruno and {Old}, Lyndsay and {Chen}, Qifeng and {Hroch}, Filip},
        title = "{Learning to denoise astronomical images with U-nets}",
      journal = {\mnras},
         year = 2021,
        month = may,
       volume = {503},
       number = {3},
        pages = {3204-3215},
          doi = {10.1093/mnras/staa3567},
archivePrefix = {arXiv},
       eprint = {2011.07002},
 primaryClass = {astro-ph.IM},
       adsurl = {https://ui.adsabs.harvard.edu/abs/2021MNRAS.503.3204V}
}

@ARTICLE{Xiang2019,
       author = {{Xiang}, Maosheng and {Ting}, Yuan-Sen and {Rix}, Hans-Walter and {Sandford}, Nathan and {Buder}, Sven and {Lind}, Karin and {Liu}, Xiao-Wei and {Shi}, Jian-Rong and {Zhang}, Hua-Wei},
        title = "{Abundance Estimates for 16 Elements in 6 Million Stars from LAMOST DR5 Low-Resolution Spectra}",
      journal = {\apjs},
         year = 2019,
        month = dec,
       volume = {245},
       number = {2},
          eid = {34},
        pages = {34},
          doi = {10.3847/1538-4365/ab5364},
archivePrefix = {arXiv},
       eprint = {1908.09727},
 primaryClass = {astro-ph.SR},
       adsurl = {https://ui.adsabs.harvard.edu/abs/2019ApJS..245...34X}
}

@ARTICLE{Xiang2022,
       author = {{Xiang}, Maosheng and {Rix}, Hans-Walter and {Ting}, Yuan-Sen and {Kudritzki}, Rolf-Peter and {Conroy}, Charlie and {Zari}, Eleonora and {Shi}, Jian-Rong and {Przybilla}, Norbert and {Ramirez-Tannus}, Maria and {Tkachenko}, Andrew and {Gebruers}, Sarah and {Liu}, Xiao-Wei},
        title = "{Stellar labels for hot stars from low-resolution spectra. I. The HotPayne method and results for 330 000 stars from LAMOST DR6}",
      journal = {\aap},
         year = 2022,
        month = jun,
       volume = {662},
          eid = {A66},
        pages = {A66},
          doi = {10.1051/0004-6361/202141570},
archivePrefix = {arXiv},
       eprint = {2108.02878},
 primaryClass = {astro-ph.SR},
       adsurl = {https://ui.adsabs.harvard.edu/abs/2022A&A...662A..66X}
}

@ARTICLE{Zhang2023_XP,
       author = {{Zhang}, Xiangyu and {Green}, Gregory M. and {Rix}, Hans-Walter},
        title = "{Parameters of 220 million stars from Gaia BP/RP spectra}",
      journal = {\mnras},
         year = 2023,
        month = sep,
       volume = {524},
       number = {2},
        pages = {1855-1884},
          doi = {10.1093/mnras/stad1941},
archivePrefix = {arXiv},
       eprint = {2303.03420},
 primaryClass = {astro-ph.SR},
       adsurl = {https://ui.adsabs.harvard.edu/abs/2023MNRAS.524.1855Z}
}

@ARTICLE{Zhang2024,
       author = {{Zhang}, Meng and {Xiang}, Maosheng and {Ting}, Yuan-Sen and {Wang}, Jiahui and {Li}, Haining and {Zou}, Hu and {Nie}, Jundan and {Mou}, Lanya and {Wu}, Tianmin and {Wu}, Yaqian and {Liu}, Jifeng},
        title = "{Determining Stellar Elemental Abundances from DESI Spectra with the Data-driven Payne}",
      journal = {\apjs},
         year = 2024,
        month = aug,
       volume = {273},
       number = {2},
          eid = {19},
        pages = {19},
          doi = {10.3847/1538-4365/ad51dd},
archivePrefix = {arXiv},
       eprint = {2402.06242},
 primaryClass = {astro-ph.GA},
       adsurl = {https://ui.adsabs.harvard.edu/abs/2024ApJS..273...19Z}
}

@ARTICLE{Zhou2022,
       author = {{Zhou}, Xingchen and {Gong}, Yan and {Meng}, Xian-Min and {Cao}, Ye and {Chen}, Xuelei and {Chen}, Zhu and {Du}, Wei and {Fu}, Liping and {Luo}, Zhijian},
        title = "{Extracting photometric redshift from galaxy flux and image data using neural networks in the CSST survey}",
      journal = {\mnras},
         year = 2022,
        month = may,
       volume = {512},
       number = {3},
        pages = {4593-4603},
          doi = {10.1093/mnras/stac786},
archivePrefix = {arXiv},
       eprint = {2112.08690},
 primaryClass = {astro-ph.CO},
       adsurl = {https://ui.adsabs.harvard.edu/abs/2022MNRAS.512.4593Z}
}

@ARTICLE{Pasquet2019,
       author = {{Pasquet}, Johanna and {Pasquet}, J{\'e}r{\^o}me and {Chaumont}, Marc and {Fouchez}, Dominique},
        title = "{PELICAN: deeP architecturE for the LIght Curve ANalysis}",
      journal = {\aap},
         year = 2019,
        month = jul,
       volume = {627},
          eid = {A21},
        pages = {A21},
          doi = {10.1051/0004-6361/201834473},
archivePrefix = {arXiv},
       eprint = {1901.01298},
 primaryClass = {astro-ph.IM},
       adsurl = {https://ui.adsabs.harvard.edu/abs/2019A&A...627A..21P}
}

@INPROCEEDINGS{He2016,
       author = {{He}, Kaiming and {Zhang}, Xiangyu and {Ren}, Shaoqing and {Sun}, Jian},
        title = "{Deep Residual Learning for Image Recognition}",
    booktitle = {2016 IEEE Conference on Computer Vision and Pattern Recognition (CVPR)},
         year = 2016,
        month = jun,
          eid = {1},
        pages = {1},
          doi = {10.1109/CVPR.2016.90},
archivePrefix = {arXiv},
       eprint = {1512.03385},
 primaryClass = {cs.CV},
       adsurl = {https://ui.adsabs.harvard.edu/abs/2016cvpr.confE...1H}
}

@ARTICLE{Touvron2023,
       author = {{Touvron}, Hugo and {Lavril}, Thibaut and {Izacard}, Gautier and {Martinet}, Xavier and {Lachaux}, Marie-Anne and {Lacroix}, Timoth{\'e}e and {Rozi{\`e}re}, Baptiste and {Goyal}, Naman and {Hambro}, Eric and {Azhar}, Faisal and {Rodriguez}, Aurelien and {Joulin}, Armand and {Grave}, Edouard and {Lample}, Guillaume},
        title = "{LLaMA: Open and Efficient Foundation Language Models}",
      journal = {arXiv e-prints},
         year = 2023,
        month = feb,
          eid = {arXiv:2302.13971},
        pages = {arXiv:2302.13971},
          doi = {10.48550/arXiv.2302.13971},
archivePrefix = {arXiv},
       eprint = {2302.13971},
 primaryClass = {cs.CL},
       adsurl = {https://ui.adsabs.harvard.edu/abs/2023arXiv230213971T}
}

@ARTICLE{OpenAI2023,
       author = {{OpenAI} and {Achiam}, Josh and {Adler}, Steven and {Agarwal}, Sandhini and {Ahmad}, Lama and {Akkaya}, Ilge and {Leoni Aleman}, Florencia and {Almeida}, Diogo and {Altenschmidt}, Janko and {Altman}, Sam and {Anadkat}, Shyamal and {Avila}, Red and {Babuschkin}, Igor and {Balaji}, Suchir and {Balcom}, Valerie and {Baltescu}, Paul and {Bao}, Haiming and {Bavarian}, Mohammad and {Belgum}, Jeff and {Bello}, Irwan and others},
        title = "{GPT-4 Technical Report}",
      journal = {arXiv e-prints},
         year = 2023,
        month = mar,
          eid = {arXiv:2303.08774},
        pages = {arXiv:2303.08774},
          doi = {10.48550/arXiv.2303.08774},
archivePrefix = {arXiv},
       eprint = {2303.08774},
 primaryClass = {cs.CL},
       adsurl = {https://ui.adsabs.harvard.edu/abs/2023arXiv230308774O}
}

@ARTICLE{Storrie-Lombardi1992,
  author = {{Storrie-Lombardi}, M.~C. and {Lahav}, O. and {Sodre}, Jr., L. and {Storrie-Lombardi}, L.~J.},
  title = "{Morphological Classification of Galaxies by Artificial Neural Networks}",
  journal = {\mnras},
  year = 1992,
  month = nov,
  volume = {259},
  pages = {8P},
  doi = {10.1093/mnras/259.1.8P},
  adsurl = {https://ui.adsabs.harvard.edu/abs/1992MNRAS.259P...8S}
}

@ARTICLE{Odewahn1992,
  author = {{Odewahn}, S.~C. and {Stockwell}, E.~B. and {Pennington}, R.~L. and {Humphreys}, R.~M. and {Zumach}, W.~A.},
  title = "{Automated Star/Galaxy Discrimination With Neural Networks}",
  journal = {\aj},
  year = 1992,
  month = jan,
  volume = {103},
  pages = {318},
  doi = {10.1086/116063},
  adsurl = {https://ui.adsabs.harvard.edu/abs/1992AJ....103..318O}
}

@ARTICLE{Collister2004,
       author = {{Collister}, Adrian A. and {Lahav}, Ofer},
        title = "{ANNz: Estimating Photometric Redshifts Using Artificial Neural Networks}",
      journal = {\pasp},
         year = 2004,
        month = apr,
       volume = {116},
       number = {818},
        pages = {345-351},
          doi = {10.1086/383254},
archivePrefix = {arXiv},
       eprint = {astro-ph/0311058},
 primaryClass = {astro-ph},
       adsurl = {https://ui.adsabs.harvard.edu/abs/2004PASP..116..345C}
}

@ARTICLE{Guo2020,
       author = {{Guo}, Jingjing and {Bai}, Xianyong and {Deng}, Yuanyong and {Liu}, Hui and {Lin}, Jiaben and {Su}, Jiangtao and {Yang}, Xiao and {Ji}, Kaifan},
        title = "{A Non-Linear Magnetic Field Calibration Method for Filter-Based Magnetographs by Multilayer Perceptron}",
      journal = {\solphys},
         year = 2020,
        month = jan,
       volume = {295},
       number = {1},
          eid = {5},
        pages = {5},
          doi = {10.1007/s11207-019-1573-9},
archivePrefix = {arXiv},
       eprint = {2002.02249},
 primaryClass = {astro-ph.IM},
       adsurl = {https://ui.adsabs.harvard.edu/abs/2020SoPh..295....5G}
}

@ARTICLE{Wong2020,
       author = {{Wong}, Kaze W.~K. and {Ng}, Ken K.~Y. and {Berti}, Emanuele},
        title = "{Gravitational-wave signal-to-noise interpolation via neural networks}",
      journal = {arXiv e-prints},
         year = 2020,
        month = jul,
          eid = {arXiv:2007.10350},
        pages = {arXiv:2007.10350},
          doi = {10.48550/arXiv.2007.10350},
archivePrefix = {arXiv},
       eprint = {2007.10350},
 primaryClass = {astro-ph.HE},
       adsurl = {https://ui.adsabs.harvard.edu/abs/2020arXiv200710350W}
}

@ARTICLE{Wang2021,
       author = {{Wang}, Ella Xi and {Nordlander}, Thomas and {Asplund}, Martin and {Amarsi}, Anish M. and {Lind}, Karin and {Zhou}, Yixiao},
        title = "{3D NLTE spectral line formation of lithium in late-type stars}",
      journal = {\mnras},
         year = 2021,
        month = jan,
       volume = {500},
       number = {2},
        pages = {2159-2176},
          doi = {10.1093/mnras/staa3381},
archivePrefix = {arXiv},
       eprint = {2010.15248},
 primaryClass = {astro-ph.SR},
       adsurl = {https://ui.adsabs.harvard.edu/abs/2021MNRAS.500.2159W}
}

@ARTICLE{Li2022_LAMOST,
       author = {{Li}, Xiangru and {Wang}, Zhu and {Zeng}, Si and {Liao}, Caixiu and {Du}, Bing and {Kong}, Xiao and {Li}, Haining},
        title = "{Estimation of Stellar Atmospheric Parameters from LAMOST DR8 Low-resolution Spectra with 20 {\ensuremath{\leq}} S/N < 30}",
      journal = {Research in Astronomy and Astrophysics},
         year = 2022,
        month = jun,
       volume = {22},
       number = {6},
          eid = {065018},
        pages = {065018},
          doi = {10.1088/1674-4527/ac65e7},
archivePrefix = {arXiv},
       eprint = {2204.06301},
 primaryClass = {astro-ph.GA},
       adsurl = {https://ui.adsabs.harvard.edu/abs/2022RAA....22f5018L}
}

@ARTICLE{Vynatheya2022,
       author = {{Vynatheya}, Pavan and {Hamers}, Adrian S. and {Mardling}, Rosemary A. and {Bellinger}, Earl P.},
        title = "{Algebraic and machine learning approach to hierarchical triple-star stability}",
      journal = {\mnras},
         year = 2022,
        month = nov,
       volume = {516},
       number = {3},
        pages = {4146-4155},
          doi = {10.1093/mnras/stac2540},
archivePrefix = {arXiv},
       eprint = {2207.03151},
 primaryClass = {astro-ph.SR},
       adsurl = {https://ui.adsabs.harvard.edu/abs/2022MNRAS.516.4146V}
}

@ARTICLE{Winecki2024,
       author = {{Winecki}, Dominik and {Kochanek}, Christopher S.},
        title = "{Photometry of Saturated Stars with Neural Networks}",
      journal = {\apj},
         year = 2024,
        month = aug,
       volume = {971},
       number = {1},
          eid = {61},
        pages = {61},
          doi = {10.3847/1538-4357/ad5a0b},
archivePrefix = {arXiv},
       eprint = {2404.15405},
 primaryClass = {astro-ph.SR},
       adsurl = {https://ui.adsabs.harvard.edu/abs/2024ApJ...971...61W}
}

@article{Tikhonov1963,
  author = {Tikhonov, A. N.},
  title = {Solution of incorrectly formulated problems and the regularization method},
  journal = {Soviet Mathematics Doklady},
  volume = {4},
  pages = {1035--1038},
  year = {1963}
}

@article{HoerlKennard1970,
  author = {Hoerl, A. E. and Kennard, R. W.},
  title = {Ridge regression: Biased estimation for nonorthogonal problems},
  journal = {Technometrics},
  volume = {12},
  number = {1},
  pages = {55--67},
  year = {1970}
}

@ARTICLE{Luo2015,
       author = {{Luo}, A. -Li and {Zhao}, Yong-Heng and {Zhao}, Gang and {Deng}, Li-Cai and {Liu}, Xiao-Wei and {Jing}, Yi-Peng and {Wang}, Gang and {Zhang}, Hao-Tong and {Shi}, Jian-Rong and {Cui}, Xiang-Qun and others},
        title = "{The first data release (DR1) of the LAMOST regular survey}",
      journal = {Research in Astronomy and Astrophysics},
         year = 2015,
        month = aug,
       volume = {15},
       number = {8},
          eid = {1095},
        pages = {1095},
          doi = {10.1088/1674-4527/15/8/002},
archivePrefix = {arXiv},
       eprint = {1505.01570},
 primaryClass = {astro-ph.GA},
       adsurl = {https://ui.adsabs.harvard.edu/abs/2015RAA....15.1095L}
}

@ARTICLE{Chen2024,
       author = {{Chen}, Boquan and {Ting}, Yuan-Sen and {Hayden}, Michael},
        title = "{The dawn is quiet here: Rise in [$\alpha$/Fe] is a signature of massive gas accretion that fueled the proto-Milky Way}",
      journal = {\pasa},
         year = 2024,
        month = oct,
       volume = {41},
          eid = {e063},
        pages = {e063},
          doi = {10.1017/pasa.2024.56},
archivePrefix = {arXiv},
       eprint = {2308.15976},
 primaryClass = {astro-ph.GA},
       adsurl = {https://ui.adsabs.harvard.edu/abs/2024PASA...41...63C}
}

@ARTICLE{Chen2025,
       author = {{Chen}, Boquan and {Orkney}, Matthew D.~A. and {Ting}, Yuan-Sen and {Hayden}, Michael R.},
        title = "{The dawn is quiet II: Gaia XP constraints on the Milky Way's proto-Galaxy from very metal-poor MDF tails}",
      journal = {arXiv e-prints},
         year = 2025,
        month = nov,
          eid = {arXiv:2511.18901},
        pages = {arXiv:2511.18901},
          doi = {10.48550/arXiv.2511.18901},
archivePrefix = {arXiv},
       eprint = {2511.18901},
 primaryClass = {astro-ph.GA},
       adsurl = {https://ui.adsabs.harvard.edu/abs/2025arXiv251118901C}
}

@ARTICLE{Andrae2023b,
       author = {{Andrae}, Ren{\'e} and {Rix}, Hans-Walter and {Chandra}, Vedant},
        title = "{Robust Data-driven Metallicities for 175 Million Stars from Gaia XP Spectra}",
      journal = {\apjs},
         year = 2023,
        month = jul,
       volume = {267},
       number = {1},
          eid = {8},
        pages = {8},
          doi = {10.3847/1538-4365/acd53e},
archivePrefix = {arXiv},
       eprint = {2303.01762},
 primaryClass = {astro-ph.SR},
       adsurl = {https://ui.adsabs.harvard.edu/abs/2023ApJS..267....8A}
}

@ARTICLE{Fallows2024,
       author = {{Fallows}, Connor P. and {Sanders}, Jason L.},
        title = "{Stellar atmospheric parameters from Gaia BP/RP spectra using uncertain neural networks}",
      journal = {\mnras},
         year = 2024,
        month = jun,
       volume = {531},
       number = {1},
        pages = {2126-2147},
          doi = {10.1093/mnras/stae1303},
archivePrefix = {arXiv},
       eprint = {2405.10699},
 primaryClass = {astro-ph.SR},
       adsurl = {https://ui.adsabs.harvard.edu/abs/2024MNRAS.531.2126F}
}

@ARTICLE{Yang2025,
       author = {{Yang}, Lin and {Yuan}, Haibo and {Huang}, Bowen and {Zhang}, Ruoyi and {Beers}, Timothy C. and {Xiao}, Kai and {Xu}, Shuai and {Huang}, Yang and {Xiang}, Maosheng and {Zhang}, Meng and {Zhang}, Jinming},
        title = "{Metallicities of 20 Million Giant Stars Based on Gaia XP Spectra}",
      journal = {\apjs},
         year = 2025,
        month = jul,
       volume = {279},
       number = {1},
          eid = {7},
        pages = {7},
          doi = {10.3847/1538-4365/add5e3},
archivePrefix = {arXiv},
       eprint = {2409.03676},
 primaryClass = {astro-ph.SR},
       adsurl = {https://ui.adsabs.harvard.edu/abs/2025ApJS..279....7Y}
}

@ARTICLE{Hattori2025,
       author = {{Hattori}, Kohei},
        title = "{Metallicity and $\alpha$-abundance for 48 Million Stars in Low-extinction Regions in the Milky Way}",
      journal = {\apj},
         year = 2025,
        month = feb,
       volume = {980},
       number = {1},
          eid = {90},
        pages = {90},
          doi = {10.3847/1538-4357/ad9686},
archivePrefix = {arXiv},
       eprint = {2407.07979},
 primaryClass = {astro-ph.GA},
       adsurl = {https://ui.adsabs.harvard.edu/abs/2025ApJ...980...90H}
}

@ARTICLE{Rix2022,
       author = {{Rix}, Hans-Walter and {Chandra}, Vedant and {Andrae}, Ren{\'e} and {Price-Whelan}, Adrian M. and {Weinberg}, David H. and {Conroy}, Charlie and {Fouesneau}, Morgan and {Hogg}, David W. and {De Angeli}, Francesca and {Naidu}, Rohan P. and {Xiang}, Maosheng and {Ruz-Mieres}, Daniela},
        title = "{The Poor Old Heart of the Milky Way}",
      journal = {\apj},
         year = 2022,
        month = dec,
       volume = {941},
       number = {1},
          eid = {45},
        pages = {45},
          doi = {10.3847/1538-4357/ac9e01},
archivePrefix = {arXiv},
       eprint = {2209.02722},
 primaryClass = {astro-ph.GA},
       adsurl = {https://ui.adsabs.harvard.edu/abs/2022ApJ...941...45R}
}

@ARTICLE{Gheller2022,
       author = {{Gheller}, C. and {Vazza}, F.},
        title = "{Convolutional deep denoising autoencoders for radio astronomical images}",
      journal = {\mnras},
         year = 2022,
        month = jan,
       volume = {509},
       number = {1},
        pages = {990-1009},
          doi = {10.1093/mnras/stab3044},
archivePrefix = {arXiv},
       eprint = {2110.08618},
 primaryClass = {astro-ph.IM},
       adsurl = {https://ui.adsabs.harvard.edu/abs/2022MNRAS.509..990G}
}

\end{document}